\RequirePackage{fix-cm}

\documentclass[twocolumn,epjc3]{svjour3}
\smartqed  
\RequirePackage{fix-cm}

\smartqed  

\RequirePackage{graphicx}
\usepackage{xcolor}
\journalname{Eur. Phys. J. C}
\usepackage{amsmath}
\usepackage{blindtext}
\usepackage{tabularx}
\usepackage{multicol}
\title{Multicols Demo}
\begin{document}

\title{New analytical model of static black hole with a dark matter halo and parametric constraints through quasiperiodic oscillations}

\author{Uktamjon Uktamov \thanksref{e1,addr1,addr2}  
\and Sanjar Shaymatov
\thanksref{e2,addr3,addr4,addr5}
\and Bobomurat Ahmedov
\thanksref{e3,addr1,addr6,addr7}
\and Chengxun Yuan
\thanksref{e4,addr1}
}

\thankstext{e1}{corresponding author: uktam.uktamov11@gmail.com}
\thankstext{e2}{sanjar@astrin.uz}
\thankstext{e3}{ ahmedov@astrin.uz}
\thankstext{e4}{corresponding author: yuancx@hit.edu.cn}
\institute{School of Physics, Harbin Institute of Technology, Harbin 150001, People’s Republic of China
\label{addr1} 
\and {Tashkent State Technical University, Tashkent 100095, Uzbekistan}\label{addr2}
\and{Institute of Fundamental and Applied Research, National Research University TIIAME, Kori Niyoziy 39, Tashkent 100000, Uzbekistan}\label{addr3}
\and{Institute for Theoretical Physics and Cosmology,
Zhejiang University of Technology, Hangzhou 310023, China}\label{addr4}
\and{Tashkent University of Applied Sciences, Gavhar Str. 1, Tashkent 100149, Uzbekistan}\label{addr5}
\and{Institute for Advanced Studies, New Uzbekistan University, Movarounnahr str. 1, Tashkent 100000, Uzbekistan}\label{addr6}
\and{Institute of Theoretical Physics, National University of Uzbekistan, Tashkent 100174, Uzbekistan}\label{addr7}
}

\date{Received: date / Accepted: date}

\maketitle

\begin{abstract}
A novel analytical Schwarzschild-like black hole (BH) solution is derived. It exhibits a static BH with a dark matter (DM) halo characterized by a Dehnen-type density profile. This solution could represent an alternative perspective on the interaction of black hole-dark matter systems, providing new insights into the fundamental properties of DM halos. We study the properties of the newly derived BH solution by examining its spacetime curvature characteristics and energy conditions, providing insights into how the DM halo influences these fundamental characteristics. Additionally, we analyze the timelike geodesics of test particles in the obtained BH-DM spacetime, highlighting how the presence of the novel Dehnen-type DM halo alters the gravitational dynamics and modifies particle trajectories. 
Increase of the DM halo's density $\rho_s$ and characteristic scale $r_s$ leads to an outward shift of both stable and unstable circular orbits. 
Finally, we test our model by fitting it to real data from the microquasars GRO J1655-40, GRS 1915+105, and XTE J1550-564 using a statistical  Markov Chain Monte Carlo (MCMC) method. This allows us to find the best estimates for the properties of the DM halo surrounding these systems.

\end{abstract}

\section{Introduction}
In general relativity (GR), black holes (BHs) arise naturally as exact solutions to Einstein's gravitational field equations. These theoretical predictions were made a long time back; nearly a century later, observational evidence for the existence of BHs was found through modern revolutionary Event Horizon Telescope (EHT) and LIGO-VIRGO observations \cite{Abbott16a,Abbott16b,Akiyama19L1,Akiyama22L12}. These triumphal observations are expected to provide stringent tests that probe the nature and unknown aspects of BHs. However, open questions remain regarding the behavior of BHs and their interaction with their surroundings. Among them, the nature and existence of dark matter are long-standing problems in GR. Consequently, identifying evidence for dark matter (DM) within the BH environment has become a crucial task in GR. 

In many astrophysical scenarios, supermassive BHs at the centers of galaxies are believed to be enveloped by matter distributions, including DM halos. It is well established that BHs are characterized by complex and dynamic environments, and there is substantial evidence to suggest that supermassive BHs at the center of most galaxies are the primary drivers of active galactic nuclei \cite{Rees84ARAA,Kormendy95ARAA} and are surrounded by a DM halo \cite{Iocco15NatPhy,Bertone18Nature}. In addition to the possibility of DM enveloping supermassive BHs, considerable evidence suggests that DM significantly influences galactic rotation curves \cite{Rubin70ApJ}, bullet clusters \cite{Corbelli00MNRAS}, and the large-scale structure of the universe \cite{Davis85ApJ}. 

The discovery of unexpectedly flat galactic rotation curves in large spiral and elliptical galaxies led to the postulation of the existence of DM. This DM, now supported by extensive astrophysical observations, is estimated to constitute approximately 90\% of the mass of a galaxy, the remaining 10\% being luminous baryonic matter \cite{Persic96}. Initially concentrated near galactic centers, facilitating star formation and clustering, dark matter gradually migrated outward during galactic evolution, forming DM halos around the galaxy through various dynamical processes. Astrophysical observations reveal that many large spiral and elliptical galaxies harbor a central supermassive BH (sometimes binary systems) embedded within this DM halo \cite{Valluri04ApJ,Akiyama19L1,Akiyama19L6,Akiyama22L12}. A substantial amount of DM in our universe is necessary to provide a satisfactory explanation for the observed phenomena. This is supported by cosmic microwave background observations, which reveal that DM comprises roughly 27\% of the universe, alongside 68\% dark energy. The rest of the universe, only around $5 \%$, is considered normal matter. Numerous theoretical particle candidates, predicted by extensions of the standard-model, are being investigated as potential components of DM \cite{Boehm04NPB,Bertone05PhR,Feng09JCAP,Schumann19}, such as weakly interacting massive particles (WIMPs), axions, and sterile neutrinos. With this in view, properties of DM can then be probed gravitationally in the case of weak interaction within standard-model particles. Therefore, DM may be highly concentrated around supermassive BHs, commonly found in galactic centers \cite{Iocco15NatPhy,Bertone18Nature}, significantly influencing extreme \cite{Babak17PRD} and intermediate \cite{Brown07PRL,Amaro-Seoane18PRD} mass ratio inspirals and potentially revealing the distribution of the DM halo. The influence of DM halos on supermassive BHs is a critical area of research, driven by the need to understand the DM-BH interaction. DM halo structures demonstrably affect galactic rotation curves \cite{Rubin70ApJ,Bertone18Nature,Corbelli00MNRAS} and dynamics observed in events such as the Bullet Cluster collision \cite{Clowe06ApJL}. Although current models primarily describe DM as interacting gravitationally, a substantial body of evidence convincingly supports its existence \cite{Bertone05,deSwart17Nat,Wechsler18}. 

Motivated by these findings, it is essential to analyze the DM-BH interaction and explore DM models to enhance our understanding of the fundamental nature of DM. Given the fundamental importance of DM halos and their interaction with BH systems, the exploration and analysis of DM models can yield valuable insights, relying on astrophysical data and simulation studies. Motivated by this, several analytical models have been developed to describe BH solutions within DM halos, including the Einasto \cite{Merritt06ApJ,Dutton_2014}, Navarro-Frenk-White \cite{Navarro96ApJ}, and Burkert \cite{Burkert95ApJ} models, as well as the Dehnen model \cite{Dehnen93,Shukirgaliyev21A&A,Gohain24DM,Pantig22JCAP,Al-Badawi25JCAP}. Additionally, some models incorporate a DM profile linked to a phantom scalar field \cite{Li-Yang12,Shaymatov21d,Shaymatov21pdu,Shaymatov22a}. Furthermore, analytical models exhibiting supermassive BH solutions embedded in a DM halo have also been investigated in Refs. \cite{Cardoso22DM,Hou18-dm,Shen24PLB}. 

Investigations within the Dehnen-type DM halo model have also explored DM-BH interactions from various perspectives. For example, one study examined the influence of density profile slopes on the survival of low star formation efficiency star clusters after rapid gas expulsion \cite{Shukirgaliyev21A&A}. Other scenarios have also considered star clusters with Plummer and Dehnen profiles, focusing on differing slope cusps at the time of formation. Furthermore, the BH-Dehnen-type DM halo solution has been proposed to investigate ultra-faint dwarf galaxies \cite{Pantig22JCAP}. Recent studies have also introduced new BH solutions surrounded by Dehnen-type DM halos, specifically 
\cite{Gohain24DM} and 
\cite{Al-Badawi25JCAP} profiles, using Schwarzschild BH within the halo environment. Thermodynamics and null geodesics of the effective BH-DM halo system were analyzed \cite{Gohain24DM}, with a subsequent study constraining the DM halo parameters \cite{Xamidov25PDU}. Finally, the impact of the DM halo on quasinormal modes, the photon sphere radius and the BH shadow \cite{Al-Badawi25CPC,Al-Badawi25CTP_DM}, as well as gravitational waveforms from periodic orbits \cite{Alloqulov-Xamidov25}, were investigated within the solution of the BH-Dehnen-type DM halo. 

In this paper, we derive a new Schwarzschild-like BH solution immersed in a DM halo characterized by a Dehnen-type density profile $(1, 4, 2)$. To accomplish this, we adopt the method developed in~\cite{Matos05,Xu18JCAP,Al-Badawi25JCAP}. We first analyze the spacetime curvature invariants and subsequently investigate the associated energy conditions. Finally, we investigate the impact of the DM profile on timelike particle geodesics, offering deeper insights into how the DM halo influences particle geodesics.

The structure of this paper is organized as follows: In Sec.~\ref{Sec:II}, we present a new Schwarzschild-like BH solution surrounded by a DM halo with a Dehnen-type density profile. In Sec.~\ref{Sec:III}, we investigate the spacetime curvature invariants and analyze the corresponding energy conditions. In Sec.~\ref{Sec:IV}, we study the timelike particle geodesics near the black hole, focusing on the influence of the Dehnen-type DM halo with density profile $(1, 4, 2)$. Also, In Section \ref{Sec.V}, we use a statistical method (MCMC) on  QPO data from three microquasars to determine the best-fit parameters for a rotating BH  and its surrounding DM halo. Finally, we summarize our findings and conclusions in Sec.~\ref{Sec:conclusion}.

\section{Schwarzchild-like black hole spacetime with the dark matter halo}\label{Sec:II}

\begin{figure*}[ht!]\centering
\includegraphics[width=0.45\textwidth]{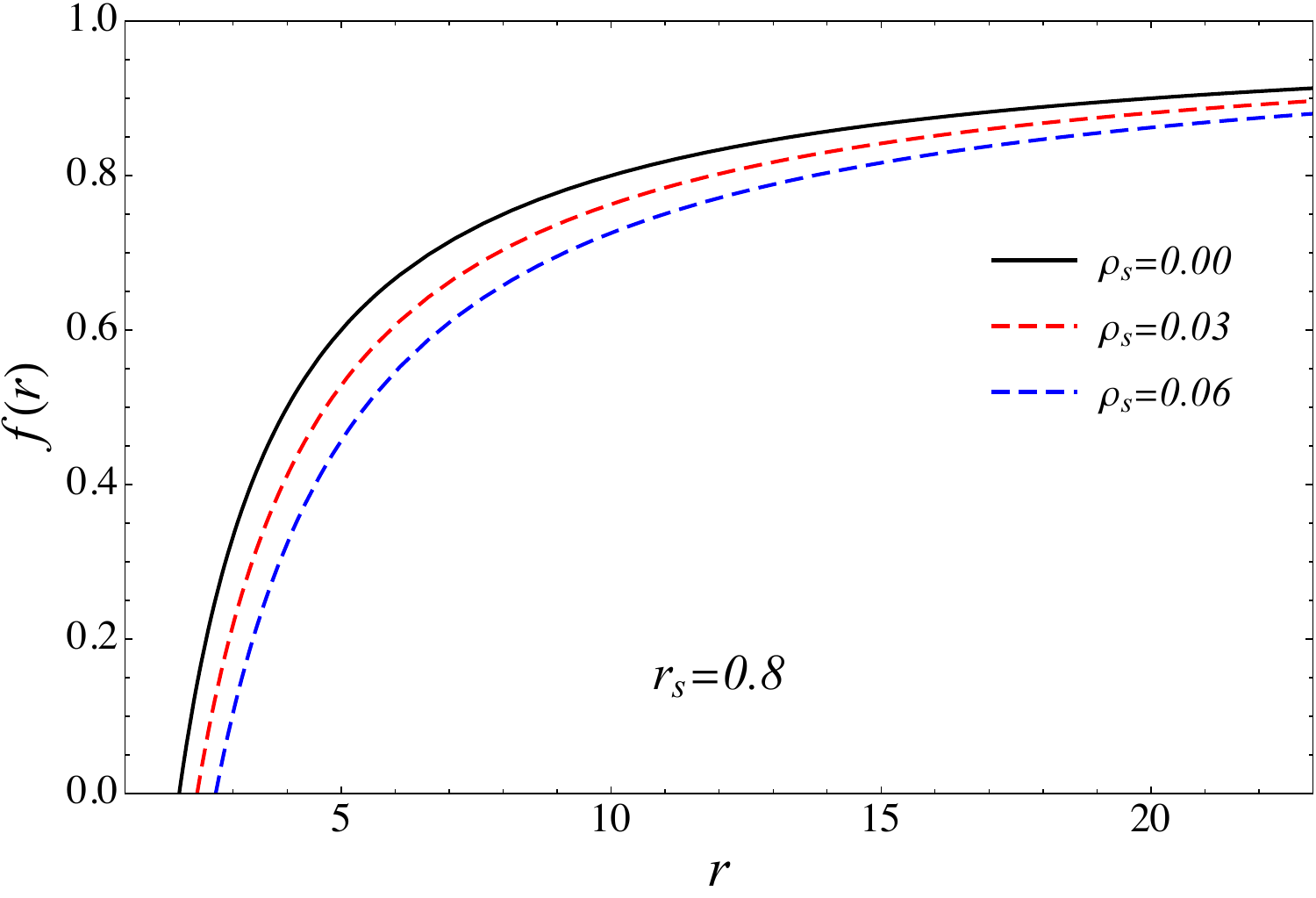}
\includegraphics[width=0.45\textwidth]{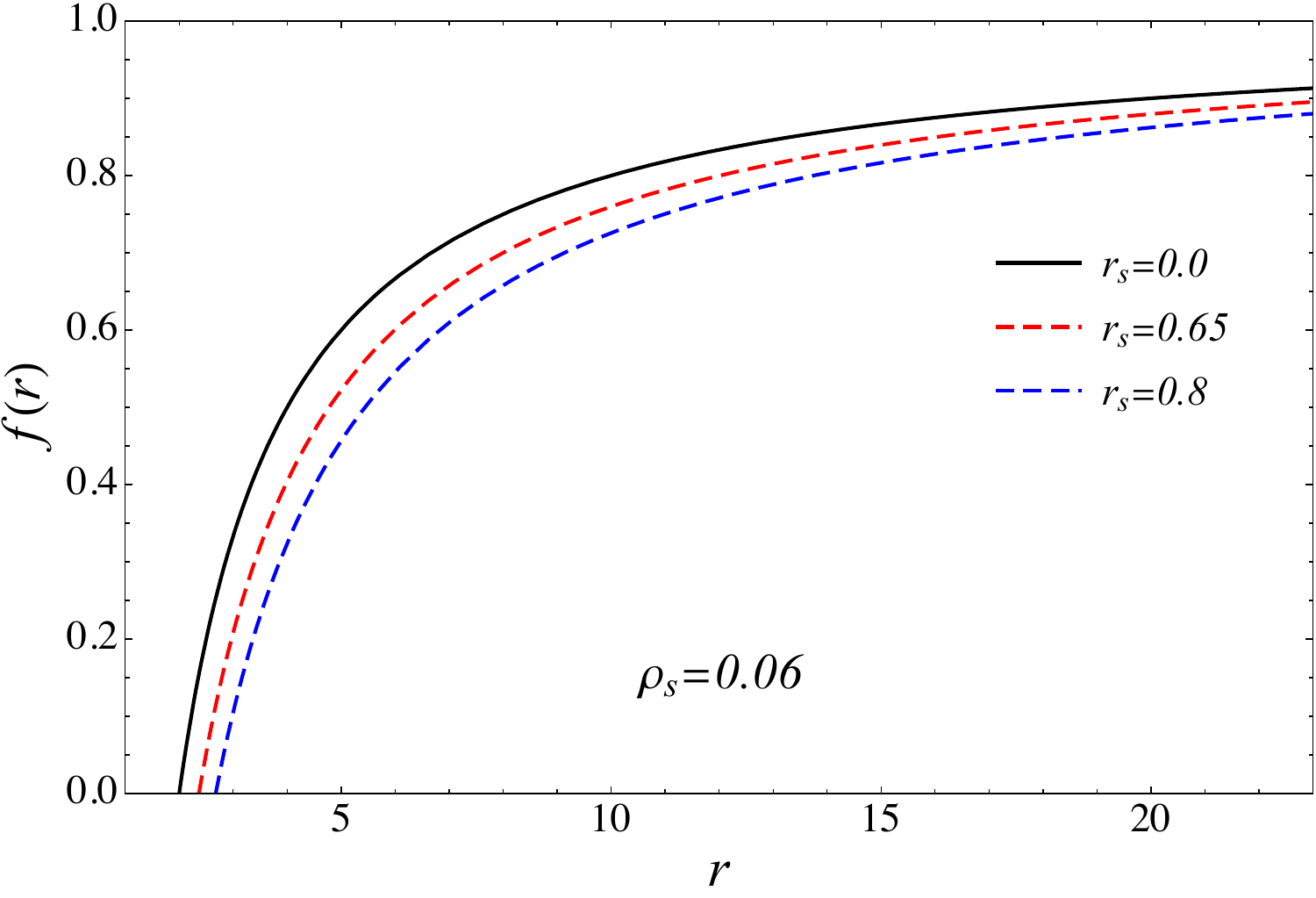}
\caption{Radial profile of metric function $f(r)$ for various values of the characteristic density $\rho_s$ (left panel) and characteristic scale $r_s$ of the DM halo (right panel). Hereafter, we shall set $M=1$, for simplicity.
\label{Fig.f(r)}}
\end{figure*}
\begin{figure*}[ht!]\centering
\centering
\includegraphics[width=0.45\textwidth]{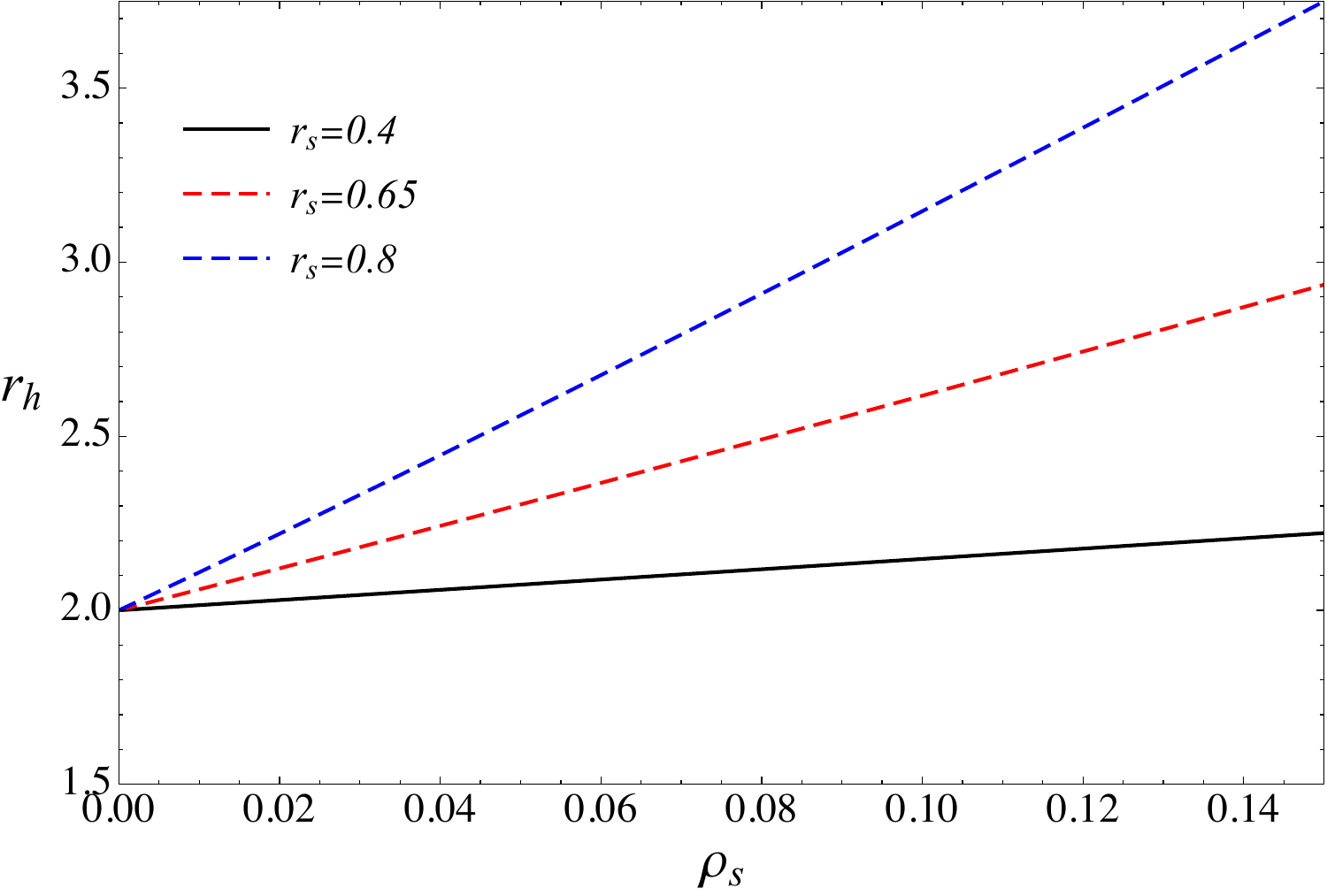}
\includegraphics[width=0.45\textwidth]{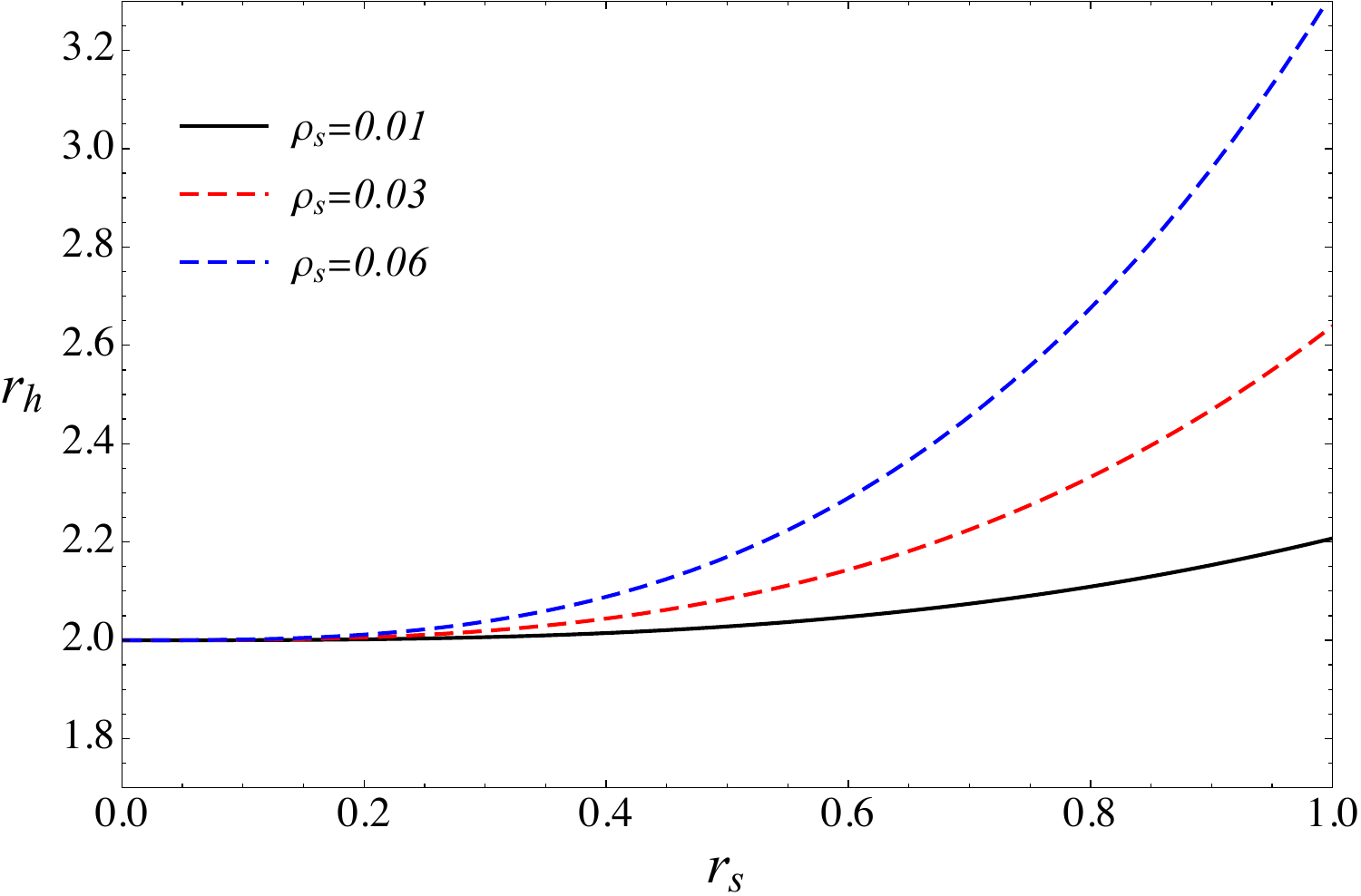}
\caption{The dependence of BH event horizon on the density $\rho_s$ (left panel) and the characteristic scale $r_s$ of the DM halo (right panel).
\label{Fig.rh}}
\end{figure*}
Here, 
we consider a Schwarzschild-like BH spacetime surrounded by a DM halo characterized by a Dehnen-type density profile. {We first consider a background spacetime involving the Dehnen-type DM distribution in the form Eq.(\ref{eq.density}) and then derive the BH-DM solution by solving Einstein's field equations.} In this case, a Schwarzschild BH can be considered with a Dehnen-type DM halo mass distribution. This is determined by the DM density distribution (see details in Ref.~\cite{Mo10book}). With this approach, we start by describing the spacetime geometry with the DM halo first and then include the Schwarzschild BH spacetime geometry. Now, we write the general form of the density profile in order to further derive the Dehnen-type DM halo mass distribution, which is defined by 
\begin{eqnarray}\label{eq.density}
    \rho(r)=\rho_s\left(\frac{r}{r_s}\right)^{-\gamma}\left[\left(\frac{r}{r_s}\right)^\alpha+1\right]^{\frac{\gamma-\beta}{\alpha}}\, ,
\end{eqnarray}
where $\rho_s$ and $r_s$ are the characteristic density and characteristic scale of the DM halo, respectively. It must be emphasized that, apart from the characteristic DM halo parameters, $\alpha$, $\beta$ and $\gamma$ are specific parameters of the density profile. For example, $\gamma$ can range only from $0$ to $3$. In this work, we use the following Dehnen-type density distribution ($\alpha,\beta,\gamma)=(1, 4, 2)$ for the DM halo. Taken together with Eq.~(\ref{eq.density}), we derive the mass profile as follows:   
\begin{eqnarray}\label{eq.mass}
    M_D(r)=\int_0^r 4\pi\rho(r_{\mathrm{_1}})r^2_{\mathrm{_1}} dr_{\mathrm{_1}}=\frac{4\pi\rho_sr_s^3}{1+\frac{r_s}{r}}\, .
\end{eqnarray}
We then define the line element of the DM halo, including redshift $A(r)$ and shape $B(r)$ functions, which is given by 
\begin{eqnarray}\label{eq.line}
    ds^2=-A(r)dt^2+\frac{dr^2}{B(r)}+r^2d\Omega^2\, ,
\end{eqnarray}
where $d\Omega^2=d\theta^2+\sin^2{\theta}d\phi^2$ is referred to as the solid angle in spherical coordinates. The point to be noted here is that $A(r)$ can be used to define the tangential velocity of particles moving within the DM halo as follows:
\begin{eqnarray}\label{eq.velocity}
v_D^2=\frac{1}{r}\partial_r\left[\log\sqrt{A(r)}\right]=\frac{M_D}{r}\, ,
\end{eqnarray}
which enables us to determine $A(r)$ as 
\begin{eqnarray}\label{eq.A(r)}
    A(r)&=&\left(1+\frac{r_s}{r}\right)^{-8\pi r_s^2\rho_s}\nonumber\\&\approx&
    1-\frac{2M_D(r)}{r_s}\left(1+\frac{r_s}{r}\right)\log{\left(1+\frac{r_s}{r}\right)}\, .
\end{eqnarray}
For the DM halo itself (\ref{eq.line}), the Einstein field equation yields as (\cite{Xu18JCAP,Al-Badawi25JCAP}):
\begin{eqnarray}\label{eq.Einstein 1}
    R_{\mu\nu}-\frac{1}{2}g_{\mu\nu}R=8\pi T_{\mu\nu}(D)\, ,
\end{eqnarray}
where the source part $T_{\mu\nu}(D)$ is given by 
\begin{eqnarray}T_\mu^\nu(D)=g^{\nu\alpha}T_{\mu\alpha}(D)=\text{diag}[-\rho(r),P_r(r),P_t(r),P_t(r)]\, ,\nonumber
\end{eqnarray} 
which stands for the energy-momentum tensor of the Dehnen-type DM halo spacetime. Similarly, the line element for the BH spacetime, including the DM halo, can be written as a result of the combined BH-DM halo system
\begin{eqnarray}\label{eq.line2}
    ds^2=-\left[A(r)+F_1(r)\right]dt^2+\frac{dr^2}{B(r)+F_2(r)}+r^2d\Omega^2\, ,
\end{eqnarray}
for which the Einstein field equation reads 
\begin{eqnarray}\label{eq.Einstein 2}
  R_{\mu\nu}-\frac{1}{2}g_{\mu\nu}R=8\pi \left[T_{\mu\nu}(D)+T_{\mu\nu}(BH)\right]\, , 
\end{eqnarray}
with the energy-momentum tensor $T_{\mu\nu}(BH)$ of the BH spacetime. For the combined BH-DM halo system, the Einstein field equations can be written as follows: 
(\ref{eq.Einstein 1},\ref{eq.Einstein 2}) and line elements (\ref{eq.line},\ref{eq.line2}) we will have:
\begin{eqnarray}\label{eq.comp.Einsteinfield1}
\Big[B(r)&+&F_2(r)\Big]\Big[\frac{1}{r}\frac{B'(r)+F_2'(r)}{B(r)+F_r(r)}+\frac{1}{r^2}\Big]\nonumber\\&=& B(r)\Big[\frac{1}{r}\frac{B'(r)}{B(r)}+\frac{1}{r^2}\Big],
\end{eqnarray}
\begin{eqnarray}\label{eq.comp.Einsteinfield2}
\Big[B(r)&+&F_2(r)\Big]\Big[\frac{1}{r}\frac{A'(r)+F_1'(r)}{A(r)+F_1(r)}+\frac{1}{r^2}\Big]\nonumber\\&=& B(r)\Big[\frac{1}{r}\frac{A'(r)}{A(r)}+\frac{1}{r^2}\Big]\, .    
\end{eqnarray}
Consequently, the above equations lead to the space-time metric, including the DM halo, which can be written as  
\begin{eqnarray} \label{m14}
ds^{2}&=& -\exp \left[ \int \frac{B(r)}{B(r)-\frac{2M}{r}}\left( \frac{1}{r}+
\frac{A^{\prime }(r)}{A(r)}\right)dr \right]dt^{2} \nonumber\\&-& A(r) dt^{2}+\frac{dr^{2}}{B(r)-\frac{2M}{r} }+r^{2}\left(
d\theta ^{2}+\sin ^{2}\theta d\phi ^{2}\right)\, ,\nonumber\\
\end{eqnarray}
It must be emphasized that one can recover $A(r) = B(r) = 1$ in the case of vanishing DM halo, the resulting integral gives $(1-\frac{2M}{r})$ that exhibits the Schwarzschild BH solution with the mass $M$. As a result, Eqs.~(\ref{eq.comp.Einsteinfield1}) and  (\ref{eq.comp.Einsteinfield2}) solve to give 
\begin{eqnarray}\label{eq.F}
    F_1(r)&=& exp\Big[\int \left(\frac{B(r)}{B(r)+F_2(r)}\Big(\frac{1}{r}+\frac{A'(r)}{A(r)}\Big)-\frac{1}{r}\right)dr\Big]\nonumber\\&&-A(r)\, ,\\
    F_2(r)&=&-\frac{2M}{r}\, ,
\end{eqnarray}
where prime $'$ denotes derivative with respect to $r$. At this stage, we shall for simplicity consider the condition $A(r)=B(r)$, leading to the resulting spacetime metric describing the Schwarzschild-like BH with the DM halo characterized by the Dehnen-type density distribution profile $(1, 4, 2)$. The resulting spacetime metric derived as the exact solution of the field equations (\ref{eq.Einstein 1}) reads as   
\begin{eqnarray}\label{eq.full-line}
    ds^2=-f(r)dt^2+\frac{1}{f(r)}dr^2+r^{2}\left(
d\theta ^{2}+\sin ^{2}\theta d\phi ^{2}\right)\, , 
\end{eqnarray}
where 
\begin{eqnarray}
    f(r)=
    1-\frac{2M}{r}-8\pi\rho_sr_s^2\log{\left(1+\frac{r_s}{r}\right)}\, .
\end{eqnarray}

{Our approach is similar with other author's works (see for example, \cite{Senjaya:2025via,Navarro:1995iw,Gohain:2024eer}), but using Eq.(\ref{eq.mass}) we can express obtained lapse function as:}
{
\begin{eqnarray}
   f(r)=1-\frac{2M}{r}-\frac{2M_D}{r_s}\left(1+\frac{r_s}{r}\right)\log{\left(1+\frac{r_s}{r}\right)}, 
\end{eqnarray}
here the effect of DM mass $M_D$ on the space-time metric is quite obvious  and our new model (1,4,2), with its steeper inner slope, will have more DM mass concentrated very close to the black hole compared to the $(1,4,\frac{1}{2}), (1,4,0)$ (\cite{Gohain:2024eer,Senjaya:2025via}) profiles. Generally, different choice of parameters $(\alpha,\beta,\gamma)$ means different slopes of the density profile, so leading different analytic model for DM-BH system.}

We now turn to studying the behavior of $f(r)$ and demonstrate its radial dependence in Fig.~\ref{Fig.f(r)}, highlighting how it changes under the DM halo parameters. As can be seen from Fig.~\ref{Fig.f(r)}, the behavior of $f(r)$ reduces to the flat spacetime at larger distances, especially at infinity. Furthermore, the curves are slightly shifted to the right to larger $r$ as the DM halo parameters $\rho_s$ and $r_s$ increase, resulting in the strength of the gravitational potential increasing. 

The spacetime Eq.~(\ref{eq.full-line}) has an event horizon, where a corotating observer's 4-velocity becomes null, resulting from setting $f(r)=0$. It is then straightforward to obtain the event horizon, $r_h$, in the form 
\begin{eqnarray}\label{eq.horizon}
   r_h=\frac{r_s}{\frac{4 \pi  \rho_sr_s^3}{M}\,  W\left(\frac{M\exp{{\Big[\frac{2 M+r_s}{8 \pi  \rho_sr_s^3}\Big]}}}{4 \pi  \rho_s  r_s^3}\right)-1}\, , 
\end{eqnarray}
where $W(z)$ is referred to as the Lambert function, defined as the principal solution of $W\exp{(W)} = z$ for arbitrary $z$. In Fig.~\ref{Fig.rh}, we show the behavior of the event horizon based on the density $\rho_s$ and the characteristic scale $r_s$ of the DM halo. From Fig.~\ref{Fig.rh}, one can easily observe that an increase in $\rho_s$ and $r_s$ leads to an increase in the event horizon. It can be seen from Eq.~(\ref{eq.mass}) that the DM halo parameters (i.e., $\rho_s$ and $r_s$) contribute to an increase in mass, enhancing the gravitational potential.
\section{The spacetime curvature characteristics and energy conditions}\label{Sec:III}

\begin{figure*}[ht!]\centering
\centering
\includegraphics[width=0.325\textwidth]{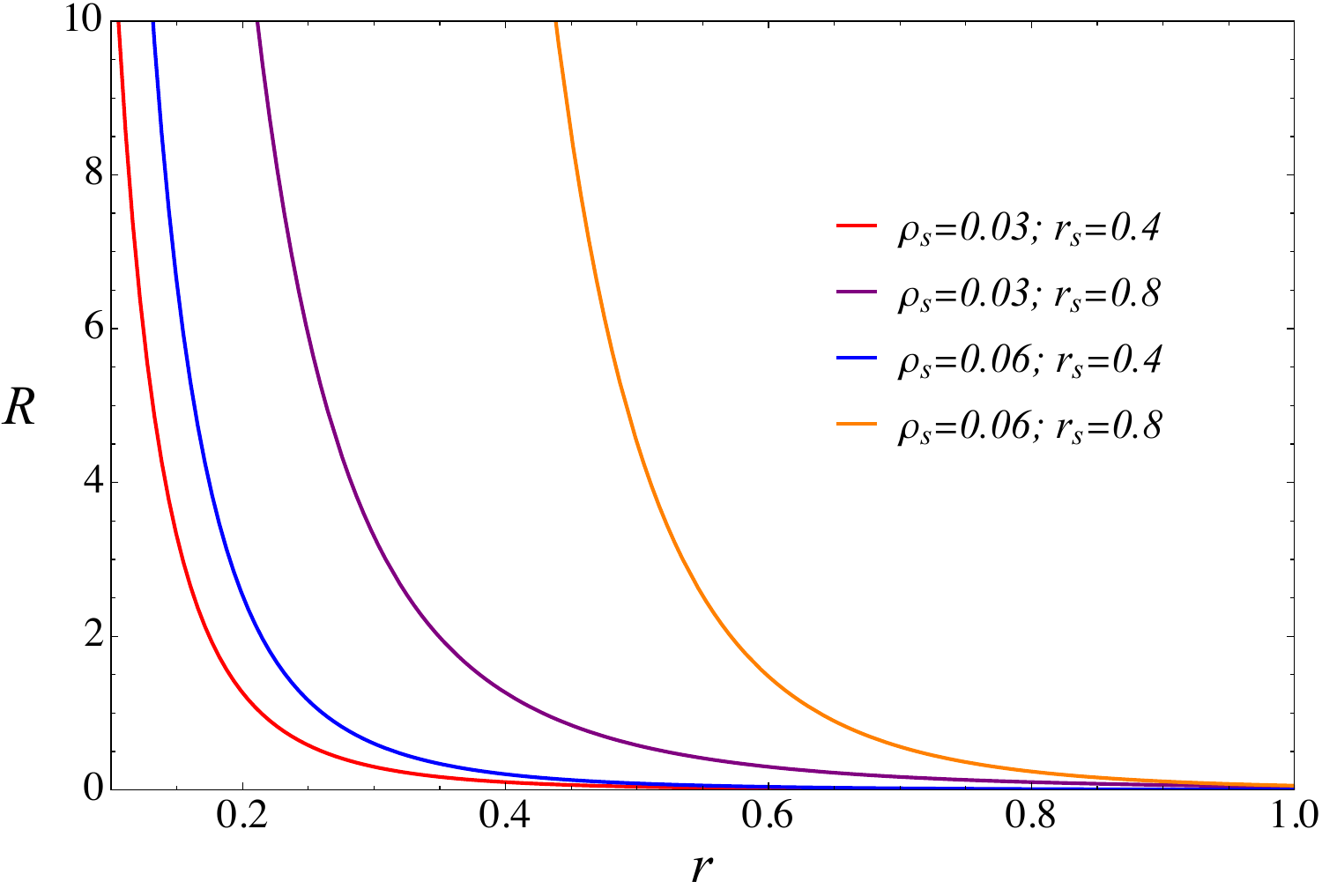}
\includegraphics[width=0.325\textwidth]{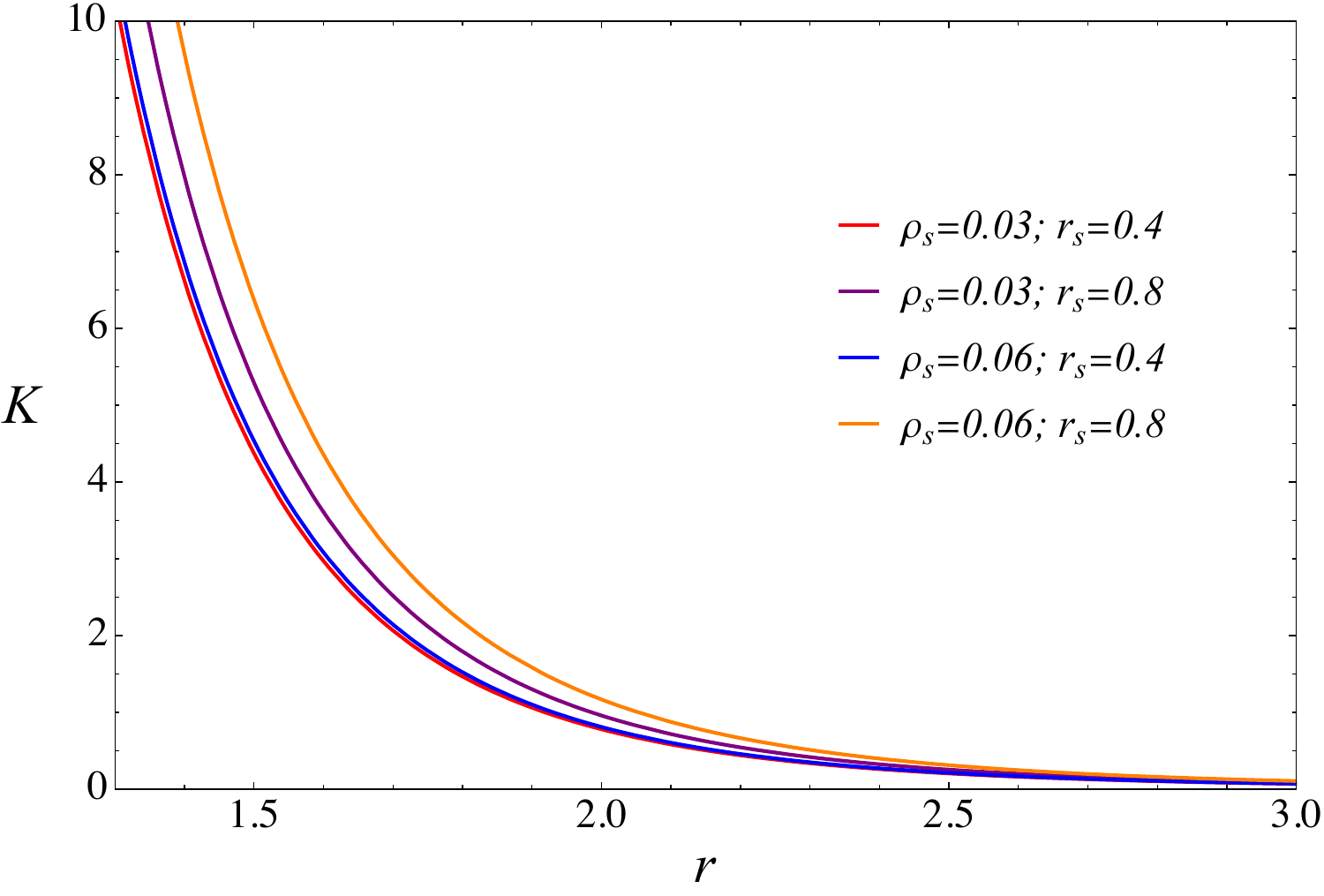}
\includegraphics[width=0.325\textwidth]{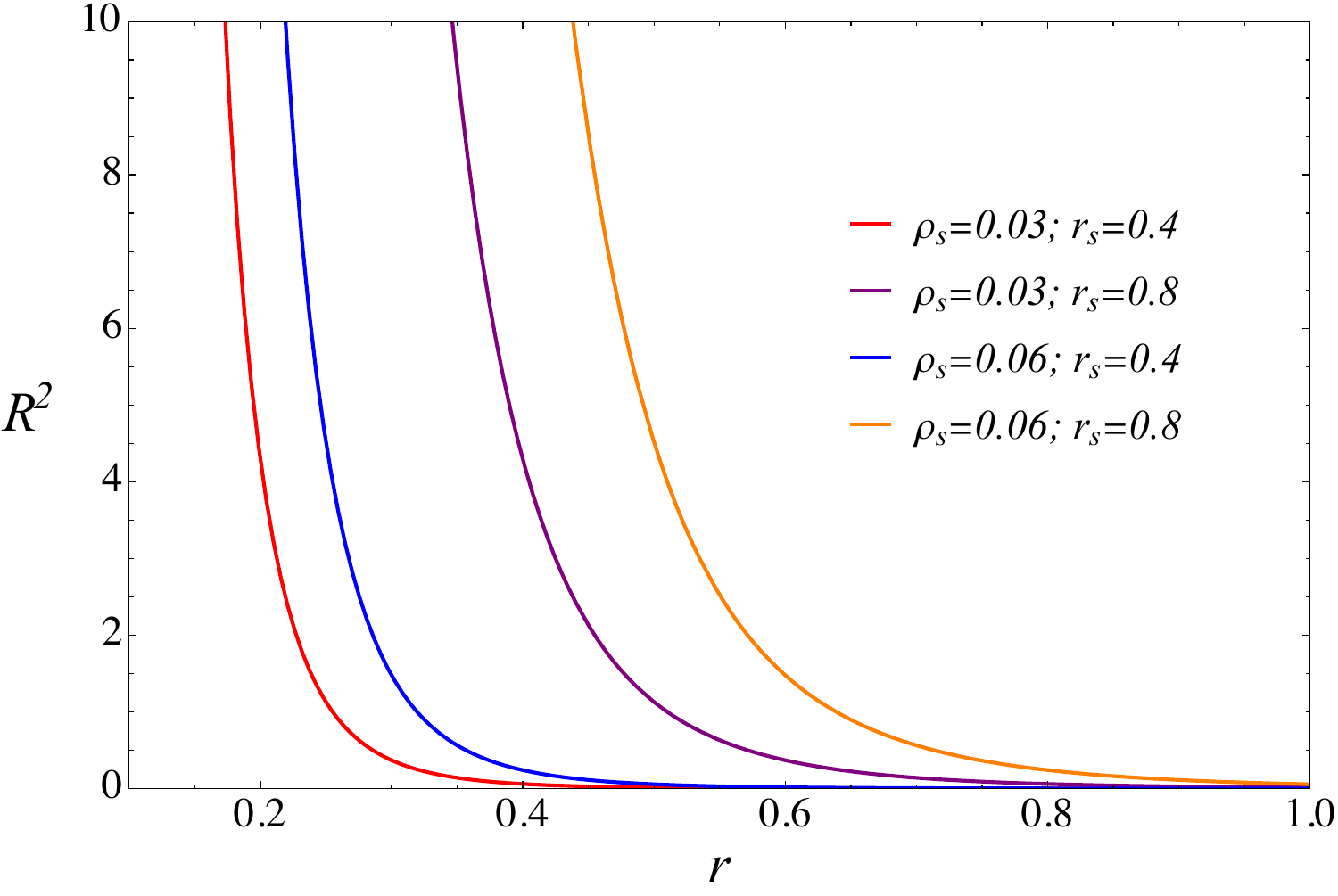}
\caption{The radial $r$ dependence of the Ricci scalar ($R$) (left panel),  the Kretschmann scalar ($K$) (middle panel), and the square of the Ricci tensor ($R^2$) (right panel) for a Schwarzschild-like BH in a DM halo for the different values of $\rho_s$ and $r_s$ parameters.
\label{Fig.Curvature}}
\end{figure*}
\begin{figure*}[ht!]\centering
\centering
\includegraphics[width=0.45\textwidth]{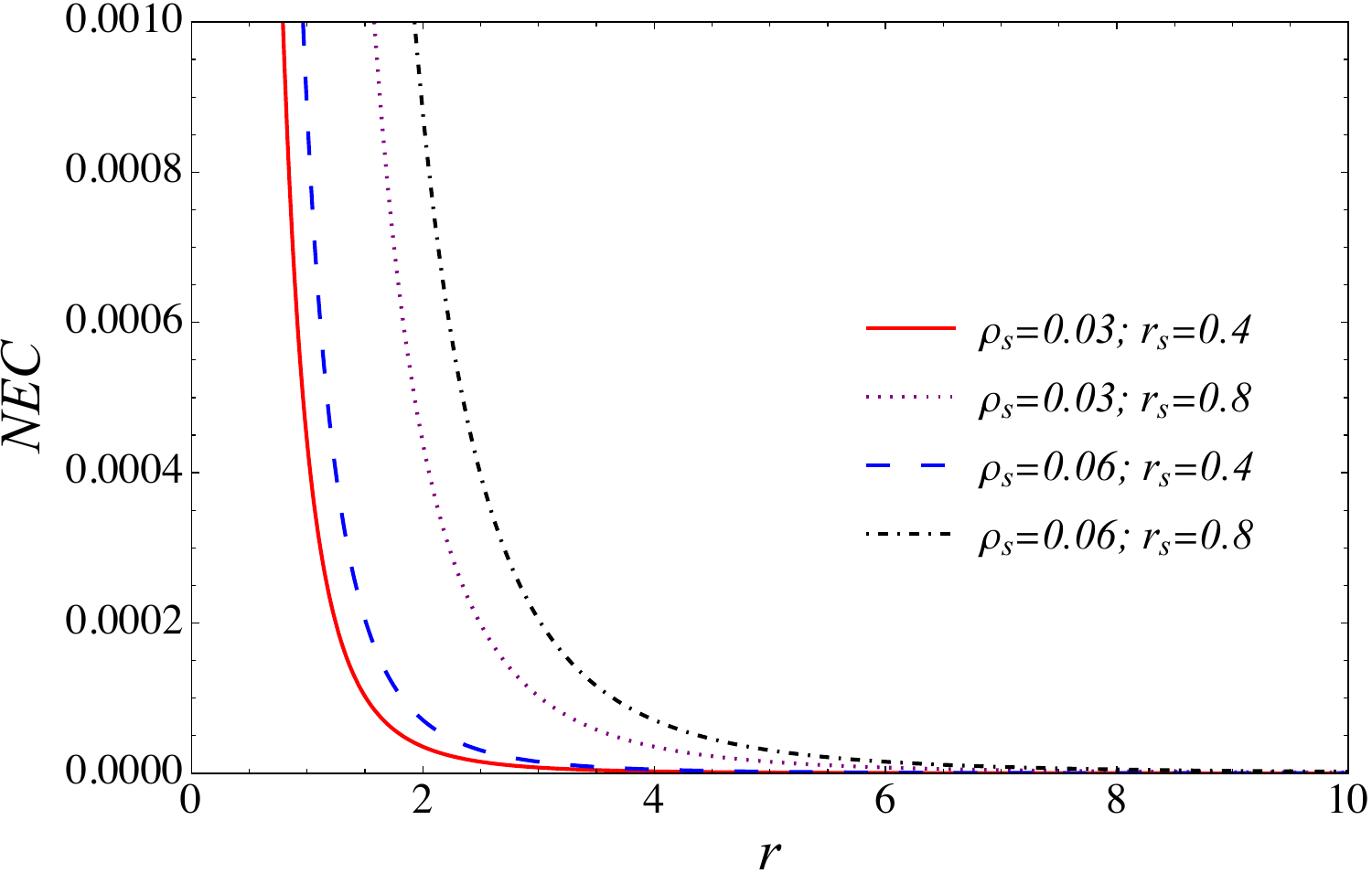}
\includegraphics[width=0.45\textwidth]{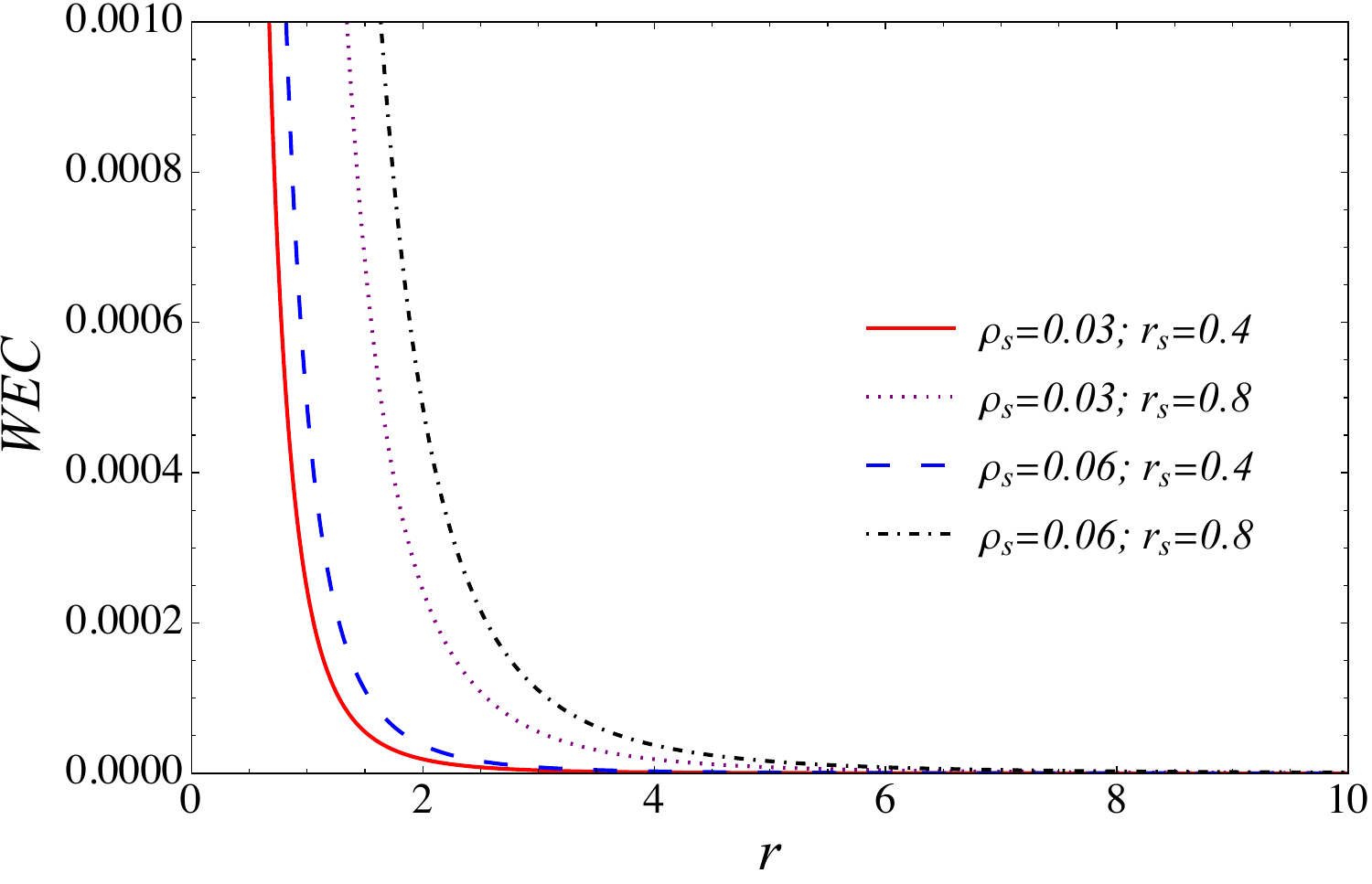}

\includegraphics[width=0.45\textwidth]{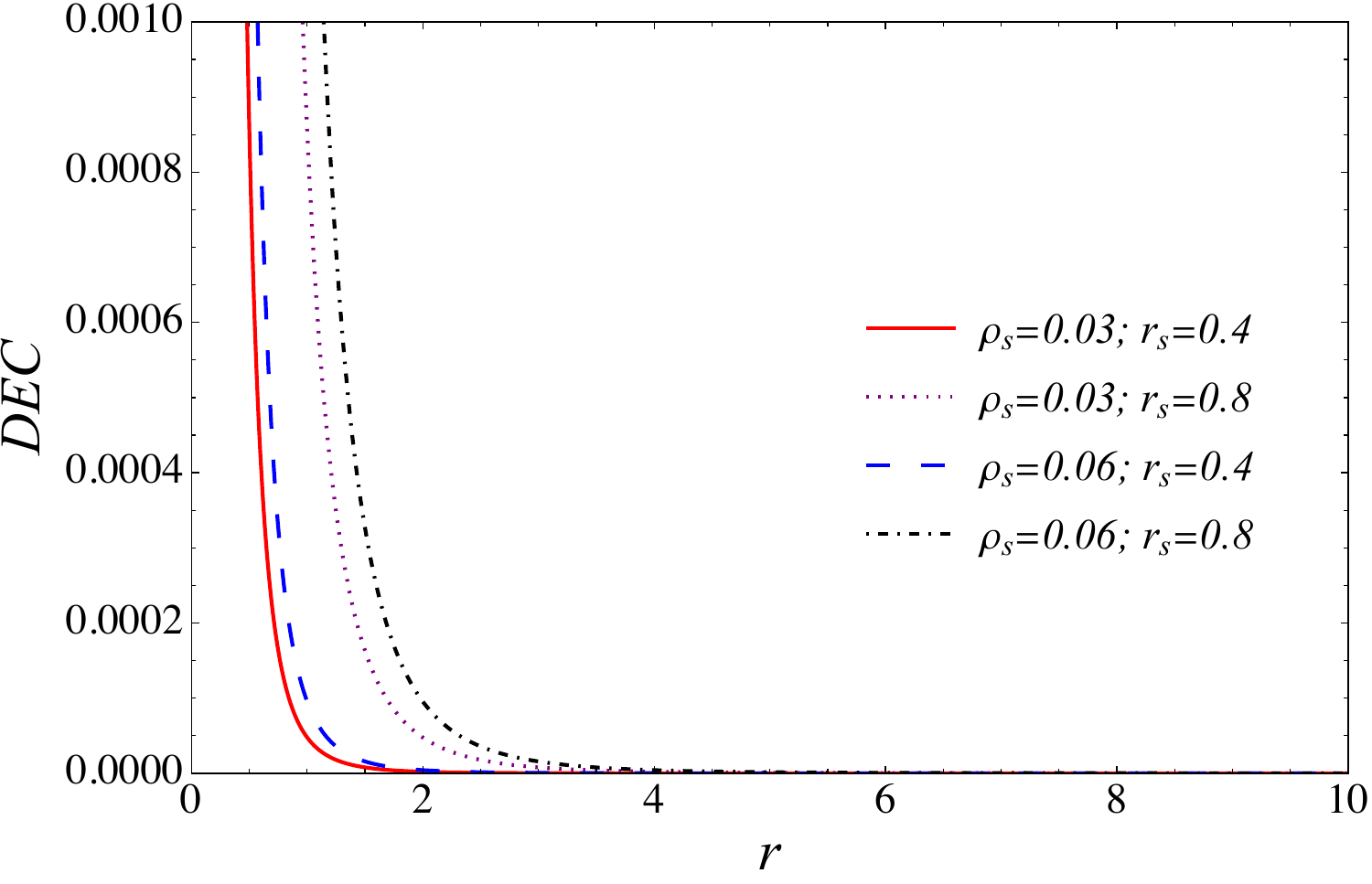}
\includegraphics[width=0.45\textwidth]{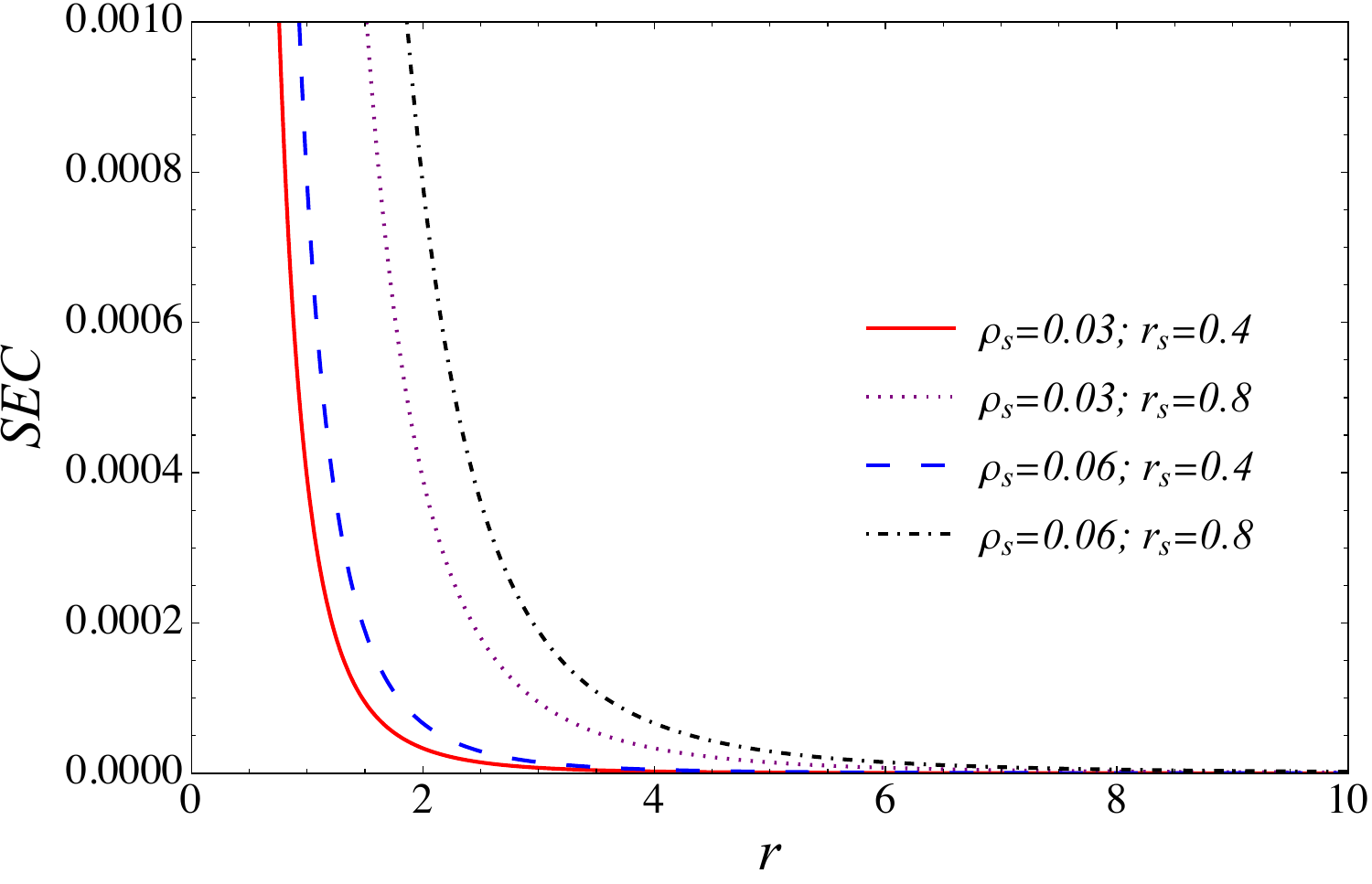}
\caption{{The radial dependence of the NEC (top left panel), WEC (top right panel), DEC (bottom left panel), SEC (bottom right panel) for various values of  
$\rho_s$ and $r_s$ parameters.
\label{Fig.EC}}}
\end{figure*}

Now we analyze the spacetime curvature invariants of a novel Schwarzschild-like BH spacetime metric with a DM halo characterized by a Dehnen-type density profile. To accomplish this, one needs to evaluate the curvature invariants of the spacetime metric, including the Ricci scalar $R$, Ricci square $R_{\mu\nu}R^{\mu\nu}$, and Kretschmann scalars $R_{\mu\nu\alpha\beta}R^{\mu\nu\alpha\beta}$. To determine whether there is a spacetime singularity or not at $r=0$, we further examine these fundamental invariants. We begin to analyze the Ricci scalar $R$, which for this BH-DM spacetime metric reads as    
\begin{eqnarray}\label{eq.Ricci}
    R=\frac{8 \pi  \rho_s  r_s^2 \left[2  \mathcal{N} \log {\mathcal{F}^{\mathcal{F}}}-r_s\mathcal{F} \mathcal{M}\right]}{r^4 \mathcal{F}}\, .
\end{eqnarray}
{One can see from Eq.(\ref{eq.Ricci}) that Ricci-flat in the Schwarzschild case (i.e., $\rho_s=0$), but not in the BH-DM system.}
 
The Kretschmann scalar reads as follows:   
\begin{eqnarray}\label{eq.Krets.}
    K&=&\frac{48 M^2}{r^6}+\frac{64 \pi  M \rho_s  r_s^2 \left[\frac{r_s (4 r+3 r_s)}{(r+r_s)^2}+2 \log {\mathcal{F}}\right]}{r^5}\nonumber\\&+&O(\rho_s^2)\, ,
\end{eqnarray}
   which reduces to the Schwarzschild case, $K=\frac{48M^2}{r^6}$, when $\rho_s=0$. Meanwhile, the Ricci square reads as 
   \begin{eqnarray}\label{eq.Ricci square}
R^2&=&\mathcal{A}\Big[\mathcal{B}\mathcal{F}^2+4 r_s^2r^2\mathcal{F}^2 \log{\mathcal{F}} \left(\mathcal{C}\log {\mathcal{F}}+\mathcal{D}\right)-4 r^2 r_s\mathcal{F}\nonumber\\ 
&\times&\mathcal{G} \log {\mathcal{F}^{\mathcal{F}}}+\mathcal{H} \log ^2{\mathcal{F}^{\mathcal{F}}}\Big]\, ,
   \end{eqnarray}
Here we note that we use the following notations in Eqs.~\eqref{eq.Ricci}-\eqref{eq.Ricci square}, which are given by 
\begin{subequations}\label{energy, momentum}
\begin{align}
   & \mathcal{A}=\frac{32 \pi ^2 \rho_s ^2r_s^4}{r^4 (r+r_s)^6},
\\
   & \mathcal{F}=1+\frac{r_s}{r},\\
   & \mathcal{B}=r_s^2 r^2\left(4 r^2+8 r r_s+5 r_s^2\right),
\\
  & \mathcal{C}=\left(r^2+2 r r_s+2 r_s^2\right), 
\\
  &  \mathcal{D}=2 r^2+4 r r_s+3 r_s^2,\\
  &\mathcal{G}=\Big[2 r^3+8 r^2 r_s+2 \left(r^3+4 r^2 r_s+5 r r_s^2+3 r_s^3\right)\\ & \times\log {\mathcal{F}}+10 r r_s^2+5 r_s^3\Big],\\
  &
  \mathcal{H}=4 r^2 \left(r^4+6 r^3 r_s+13 r^2 r_s^2+12 r r_s^3+5 r_s^4\right),\\
  &
  \mathcal{N}=\left(r^2+3 r r_s+3 r_s^2\right),
  \\
  &
  \mathcal{M}=\left[2 (r+2 r_s) \log {\mathcal{F}}+2 r+3 r_s\right]\, .
\end{align}
\end{subequations}
From these equations, one can deduce that the spacetime singularity appears at $r=0$. Furthermore, we show the radial behavior of these curvature invariants in Fig.~\ref{Fig.Curvature}. Also, the behavior of the spacetime singularity is clearly seen in Fig.~\ref{Fig.Curvature}. These curvature invariants, respectively, approach zero and infinity when $r\to \infty$ and $\to 0$, i.e.,   
\begin{eqnarray}
   \lim\limits_{r\to \infty}R,
     \vspace{3mm}R_{\mu\nu}R^{\mu\nu},
     \vspace{3mm} R_{\mu\nu\alpha\beta}R^{\mu\nu\alpha\beta}\to0\, ,
    \end{eqnarray}
and 
\begin{eqnarray}
 \lim\limits_{r\to 0} \,
    R,\, R_{\mu\nu}R^{\mu\nu}, \,
R_{\mu\nu\alpha\beta}R^{\mu\nu\alpha\beta}\to\infty\, .
\end{eqnarray}

The energy conditions are also considered to be fundamental characteristics of spacetime, like the curvature invariants. Here, we provide some insight by analyzing the corresponding energy conditions. We then turn to solving Einstein's Eq.~(\ref{eq.Einstein 1}) together with the energy-momentum tensor $$T_{\mu}^\nu=\text{diag}[-\rho(r),P_r(r),P_{\theta}(r),P_{\phi}(r)].$$
 Consequently, we derive the stress-energy tensor elements as 
{\begin{eqnarray}\label{eq.elements of the energy-mom.}
    -\rho(r)&=&P_r(r)=-\frac{\rho_sr_s^4}{r^2(r+r_s)^2}\times\\\nonumber
    &\Big[&1+\frac{r}{r_s}-(1+\frac{r}{r_s})^2\log{(1+\frac{r_s}{r})}\Big]\,,
\end{eqnarray}
\begin{eqnarray}\nonumber
    P_{\theta}(r)=P_\phi(r)=\frac{\rho_sr_s^4}{2r^2(r+r_s)^2}\, .
\end{eqnarray}
We give more explanation in Appendix \ref{Sec.App}.}
Taking these equations into account, we further examine the energy conditions for the source
fluid. It is worth noting that the stress-energy tensor must satisfy certain inequalities. Therefore, we further consider the null, weak, dominant, and strong energy conditions.
\begin{itemize}
 \item The null-energy condition (NEC) is a fundamental requirement in GR and plays a crucial role in the study of gravitational physics. It ensures that the energy density associated with matter and fields in spacetime is always positive or zero, preventing the occurrence of exotic phenomena such as negative energy densities or violations of energy conditions. The NEC is a key ingredient in the formulation of the Einstein field equations and is essential for maintaining the consistency and physical viability of the theory. Taken together, the NEC implies \cite{2017EPJC...77..542T,Al-Badawi25JCAP}:
\begin{eqnarray}\label{eq.NEC}
      T_{\mu\nu}n^\mu n^\nu\geq0,
  \end{eqnarray}
  or
  \begin{eqnarray}\nonumber
      \rho(r)+P_i(r)\geq0, \,\,\,(i=r,\theta,\phi),
  \end{eqnarray}
  where $n^\alpha$ is the null vector. One can easily conclude from Eq. (\ref{eq.elements of the energy-mom.}) that $P_r(r)+\rho(r)=0$ and:
\begin{eqnarray}\label{eq.NEC1}
    \rho(r)&+&P_{\beta}(r)\nonumber\\&=&\frac{\rho_s r_s^3 \left[ \log\left(1+\frac{r_s}{r}\right)^{2 r_s(1+\frac{r}{r_s})^2}-2 r-r_s\right]}{2 r^2 (r+r_s)^2}\, ,
  \end{eqnarray}
  with $\beta=\theta,\phi$. From Eq.(\ref{eq.NEC1}) it is obvious:
  \begin{eqnarray}\label{eq.NEC2}
      \lim_{r\to0}[\rho(r)+P_{\beta}(r)]=\infty,
  \end{eqnarray}
  \begin{eqnarray}\nonumber
      \lim_{r\to\infty}[\rho(r)+P_{\beta}(r)]=0.
  \end{eqnarray}
  The behavior of the NEC is demonstrated in Fig~\ref{Fig.EC} for various possible cases of the DM halo parameters.  
  \item The weak energy condition (WEC) is also a fundamental requirement in GR and imposes the requirement that the total energy density measured by an observer traversing a timelike curve cannot be negative, which implies \cite{2017EPJC...77..542T,Al-Badawi25JCAP}:
\begin{eqnarray}\label{eq.WEC}
\rho\geq0,\,\,\rho(r)+P_i(r)\geq0, \,\,\,(i=r,\theta,\phi).
  \end{eqnarray}
  We depict the behavior of the WEC in Fig.~\ref{Fig.EC} for various possible cases. 
  \item The dominant energy condition (DEC) states that for any future-directed causal vector field (whether timelike or null), the vector $- T^\mu_\nu \mathbf{Y}^\nu$ must also be a future-directed causal vector. This condition, which is stronger than the WEC, physically implies that the local energy density is non-negative and that energy flux cannot propagate faster than light relative to any local observer. The DEC implies following condition \cite{Al-Badawi25JCAP}:
  \begin{eqnarray}\label{eq.DEC}
      \rho(r)-|P_{\theta,\phi}|\geq0,
    \end{eqnarray}
    \begin{eqnarray}\nonumber
      \frac{\rho_sr_s^3 \left[2 (r+r_s) \log \left[\left(1+\frac{r_s}{r}\right)^{1+\frac{r}{r_s}}\right]-2 r-3 r_s\right]}{2 r^2 (r+r_s)^2}\geq0\, .
  \end{eqnarray}
  The above equation guarantees the satisfaction of this energy condition. Similar behavior is also shown in Fig.~\ref{Fig.EC} for various possible cases.
  \item Finally, the strong energy condition (SEC) can be determined by \cite{Al-Badawi25JCAP}
\begin{eqnarray}\label{eq.SEC}
      (T_{\mu\nu}-\frac{1}{2}Tg_{\mu\nu})n^\mu n^\nu\geq0,
  \end{eqnarray}
  or
  \begin{eqnarray}\nonumber
\rho+\sum_{i=1}^3P_i\geq0,
  \end{eqnarray}
  which gives 
\begin{eqnarray}\label{eq.SEC1}
      \frac{\rho_s r_s^4}{r^2(r+rs)^2}\geq0\, .
  \end{eqnarray}
\end{itemize}

On the basis of the above analysis of the energy conditions,  we obtain compelling results indicating that all energy conditions are well satisfied. This conclusion is further supported by a graphical analysis for the obtained BH spacetime metric within a Dehnen-type DM halo, presented in Eq.~\ref{eq.full-line}. Additionally, we present a clear analysis that illustrates the radial behavior of NEC, WEC, DEC, and SEC, as seen in Fig.~\ref{Fig.EC}. Both analytical and graphical investigations consistently confirm the validity of all energy conditions.     
\begin{figure*}[ht!]\centering
\includegraphics[width=1\textwidth]{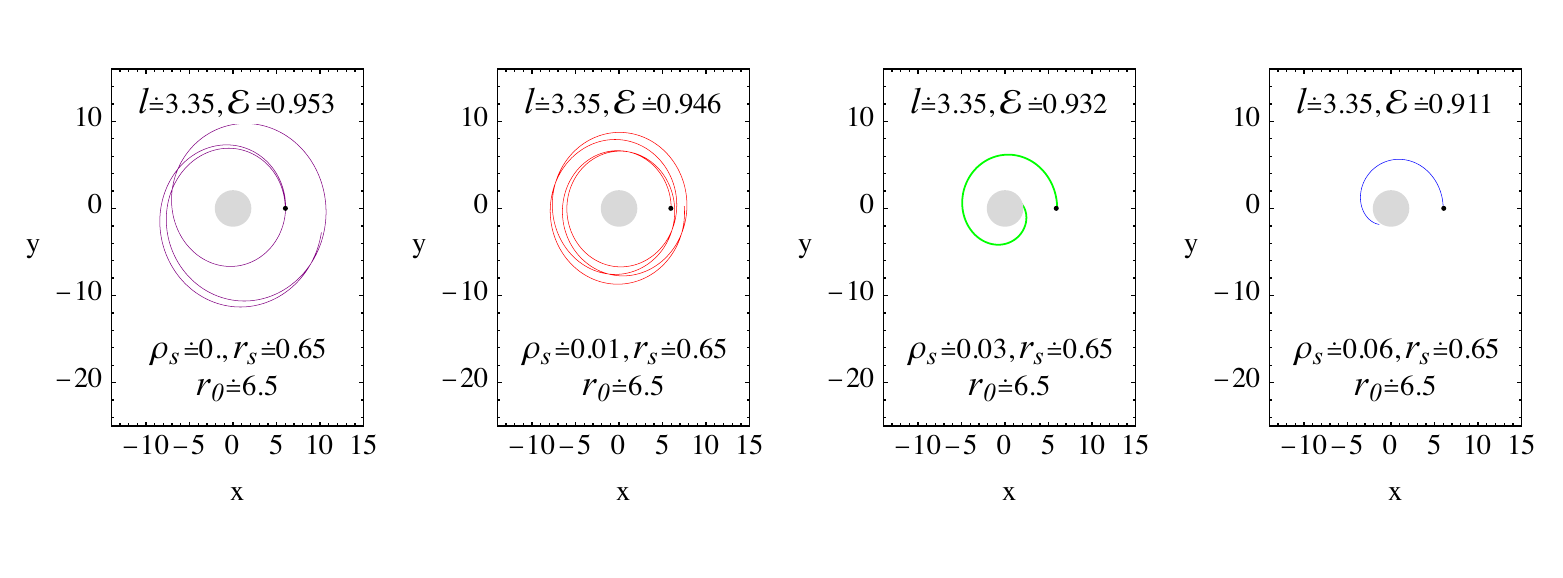}
\includegraphics[width=1\textwidth]{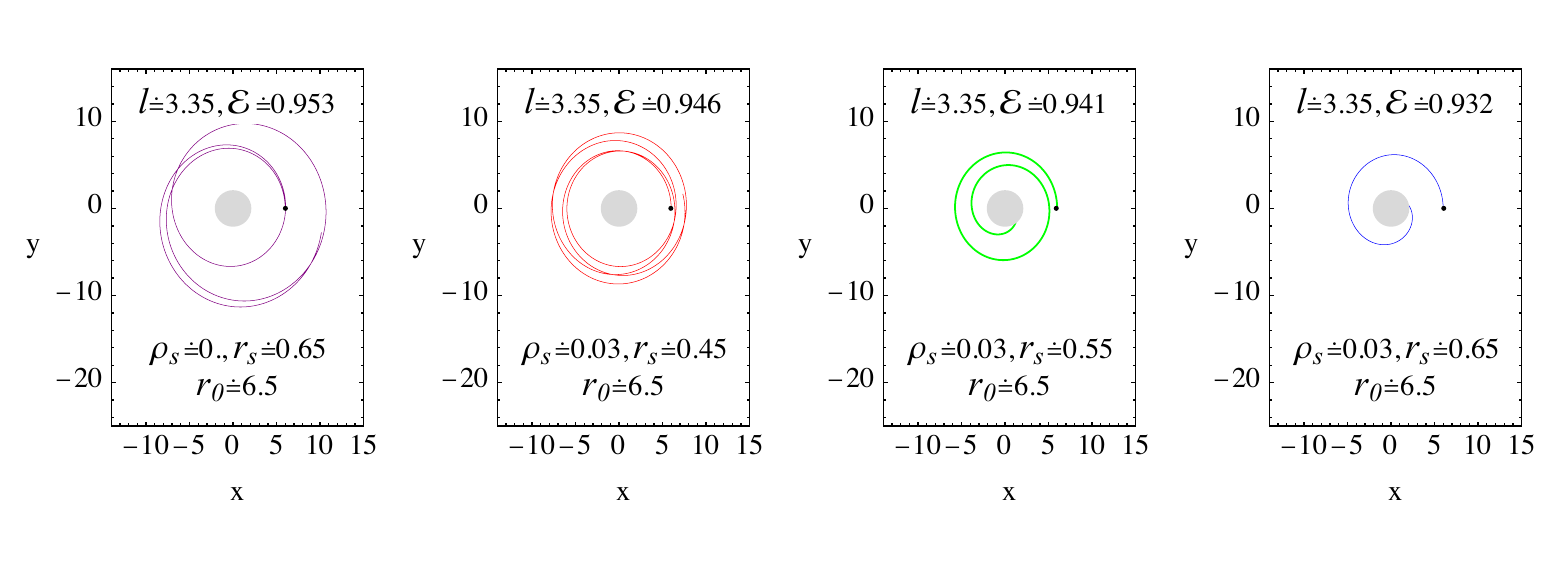}
\caption{The trajectories of the timelike particle in the vicinity of a Schwarzschild-like BH in a DM halo for varying $\rho_s$ and $r_s$ parameters. \label{Fig.trajectory}}
\end{figure*}
\section{Lyapunov exponents. Timelike geodesics}\label{Sec:IV}

\begin{figure*}[ht!]\centering
\includegraphics[width=0.45\textwidth]{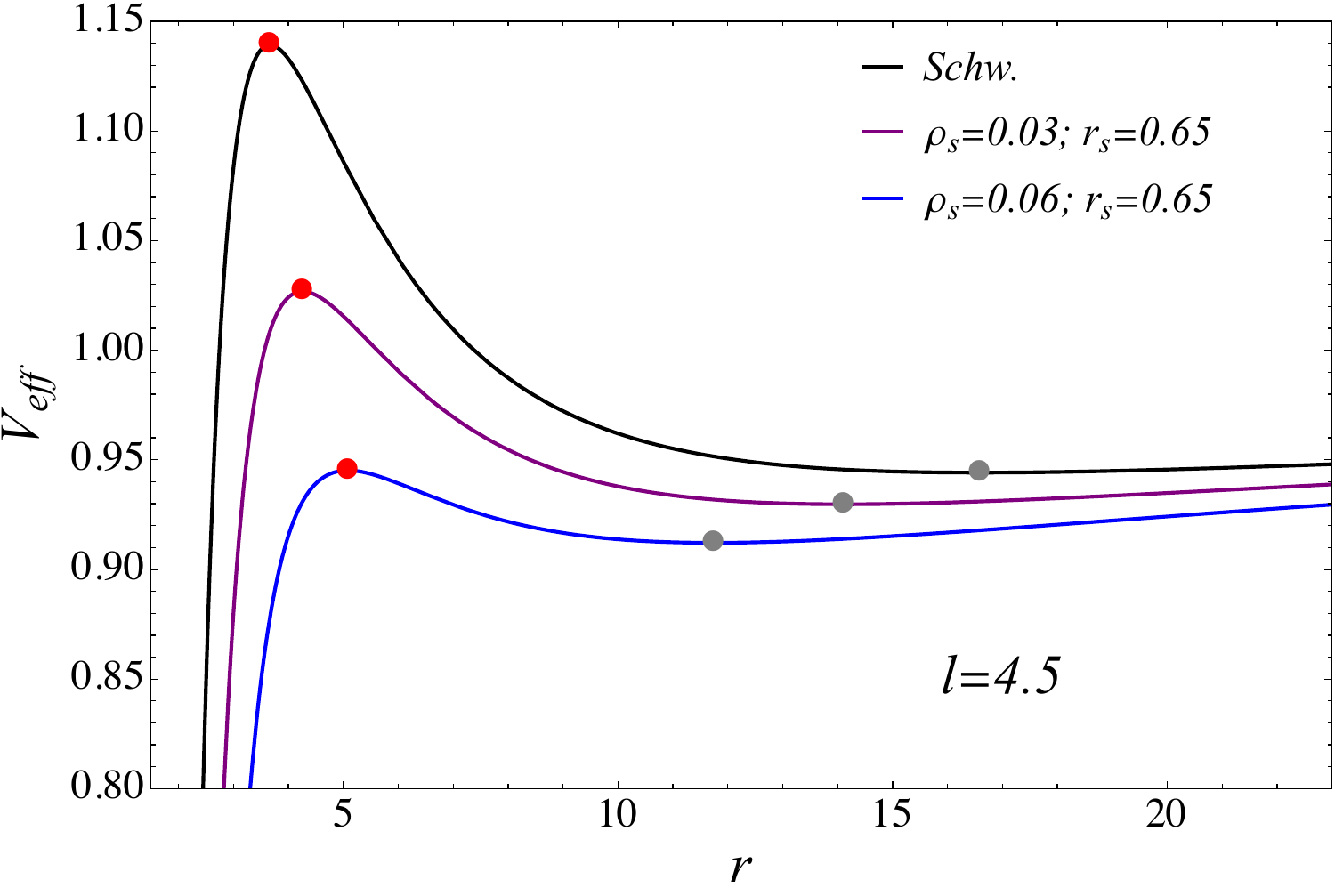}
\includegraphics[width=0.45\textwidth]{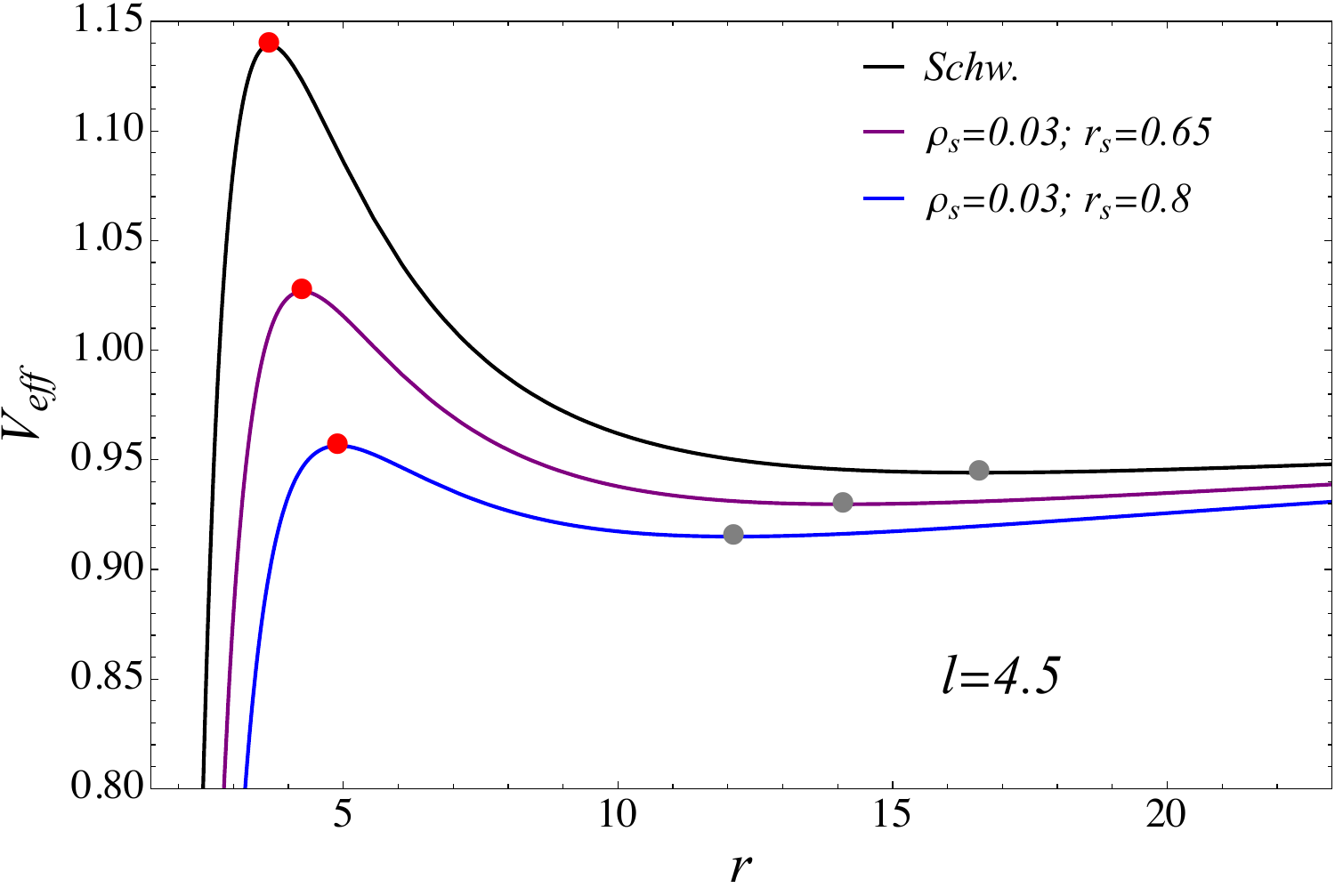}
\caption{The radial dependence of the effective potential for varying DM halo parameters, $\rho_s$ (left panel)  and $r_s$ (right panel). Here, we note that red and grey points indicate unstable and stable circular orbits, respectively. \label{Fig.Veff}}
\end{figure*}
\begin{figure*}[ht!]
\centering
\includegraphics[width=0.45\textwidth]{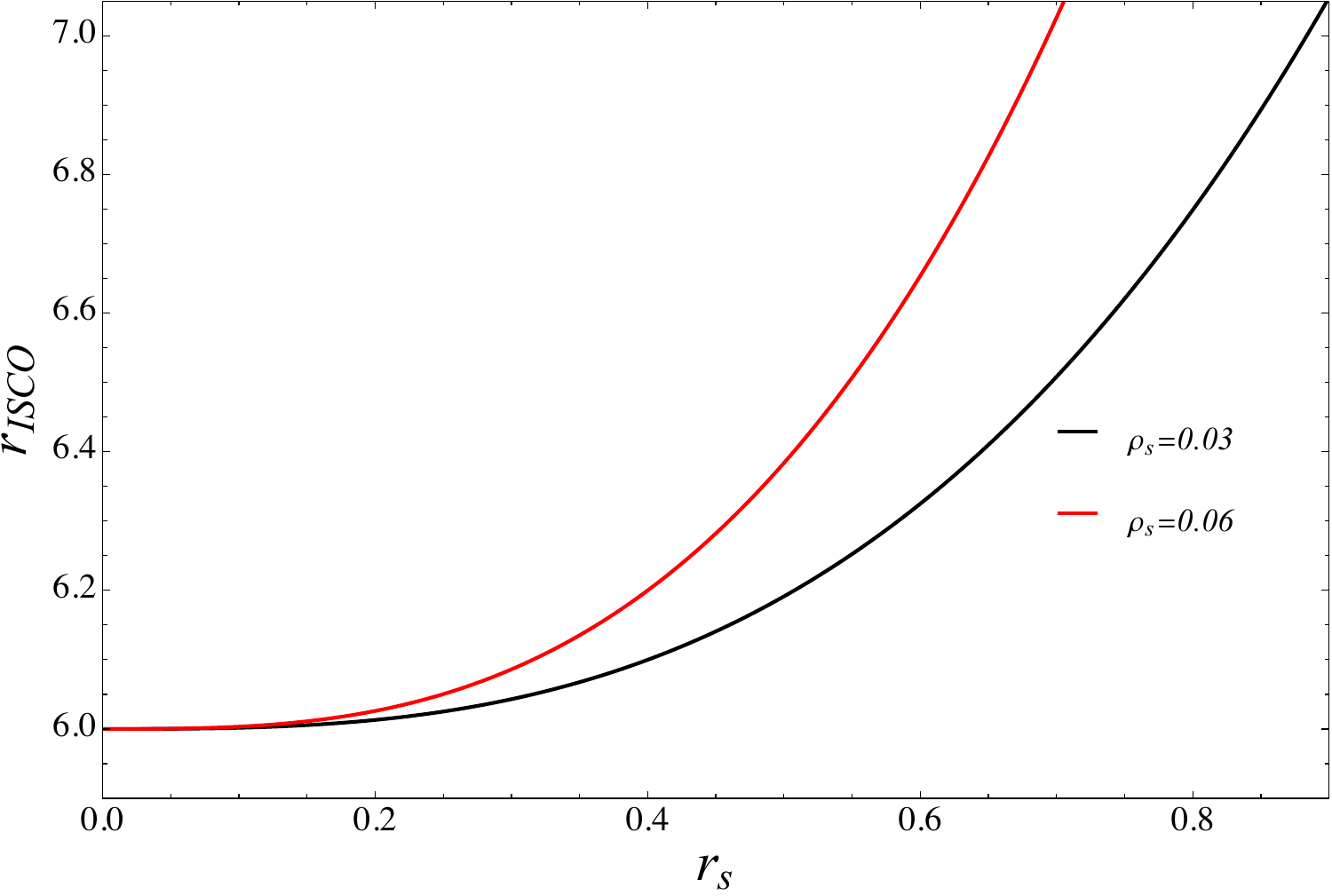}
\includegraphics[width=0.45\textwidth]{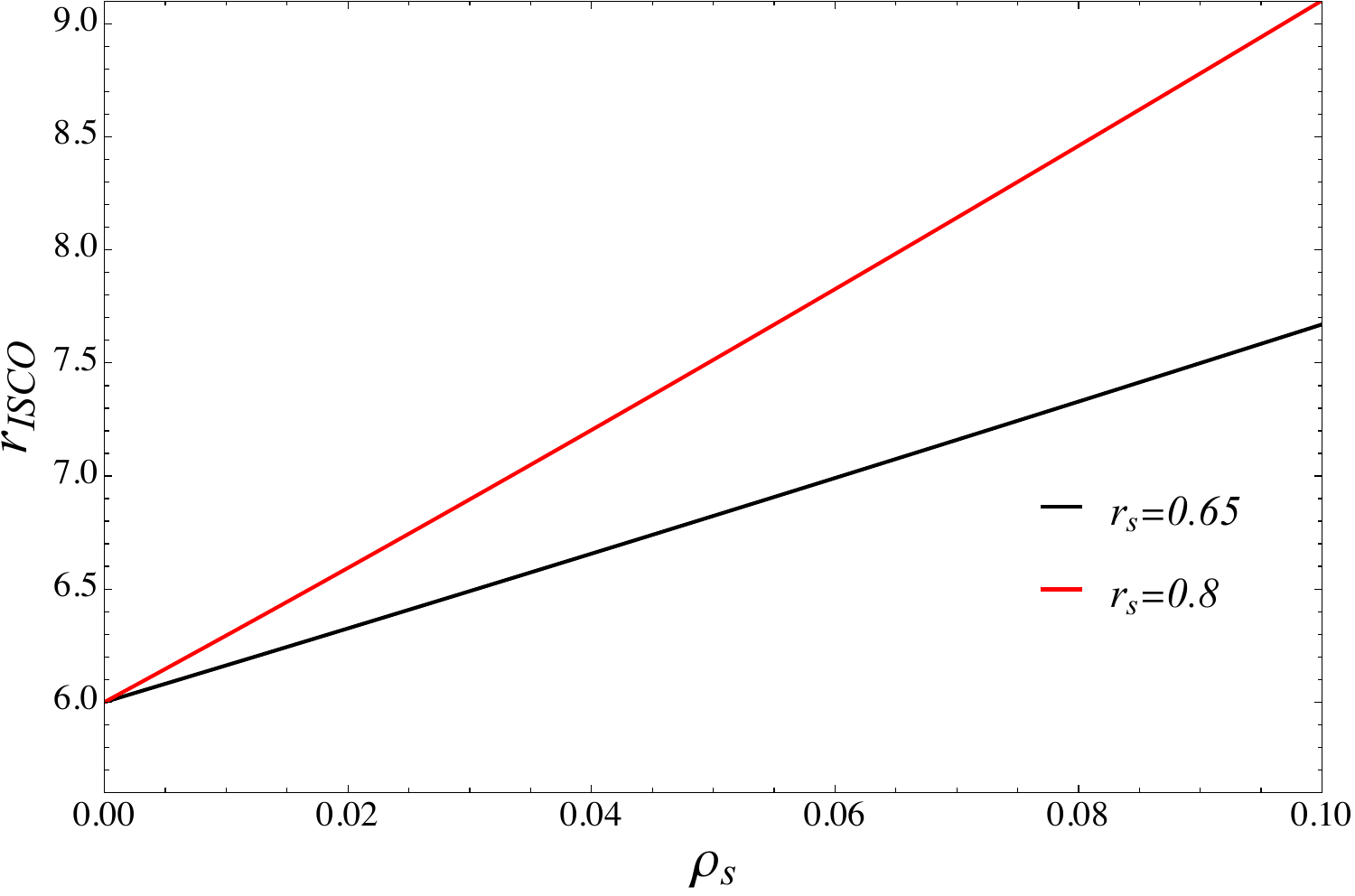}

\includegraphics[width=0.45\textwidth]{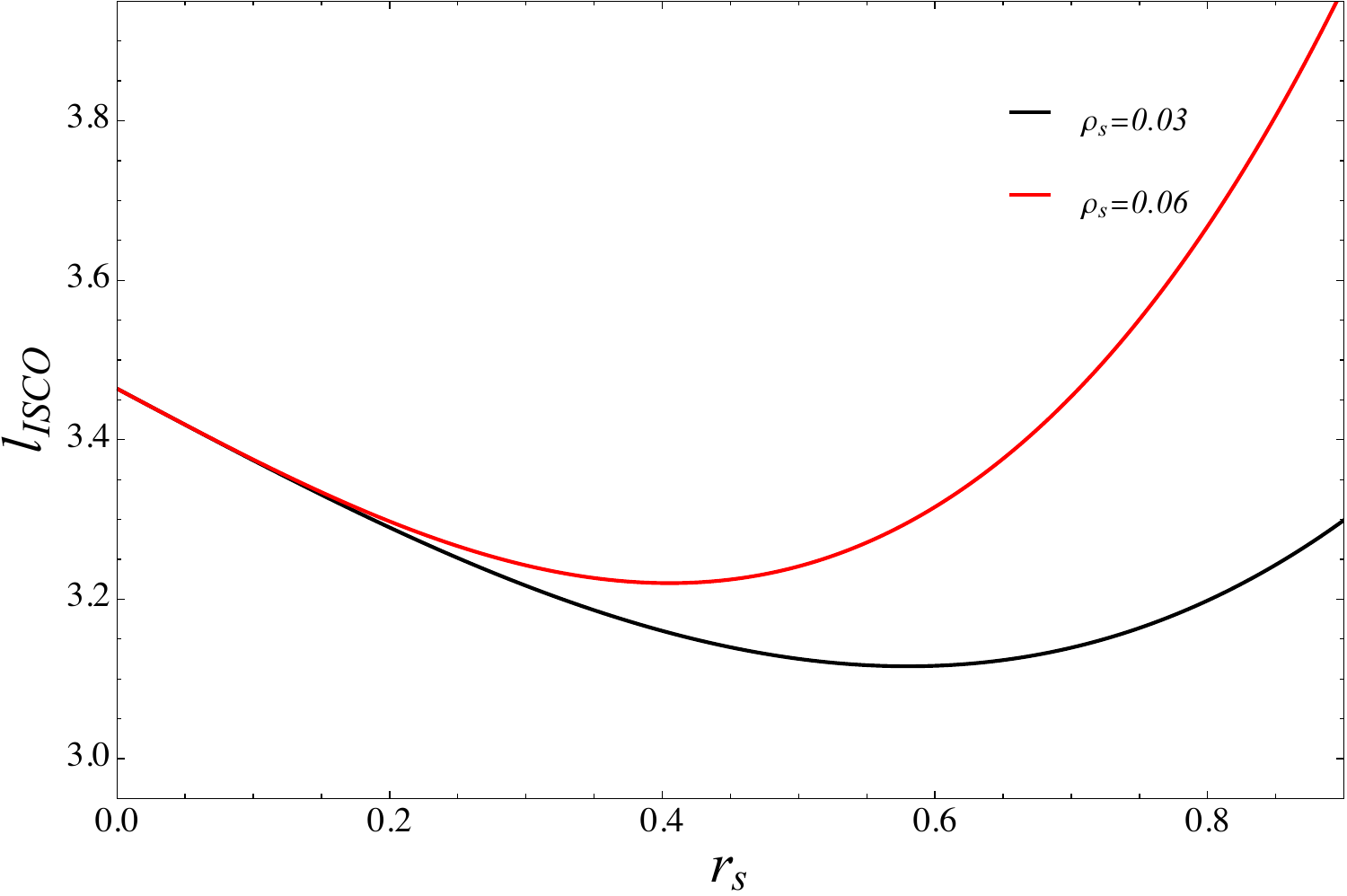}
\includegraphics[width=0.45\textwidth]{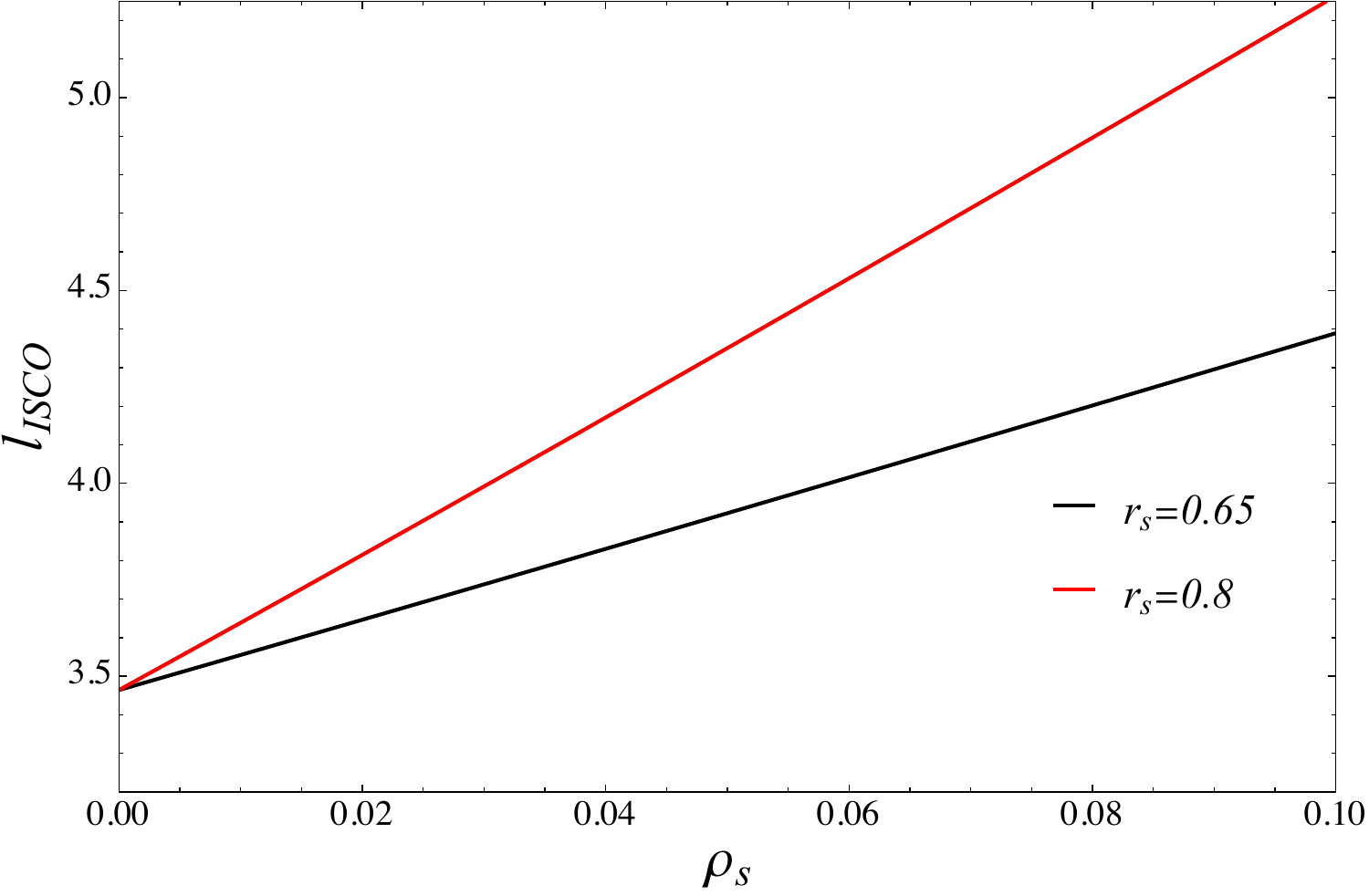}
\caption{The ISCO parameters, $r_{ISCO}$ (top row) and $l_{ISCO}$ (bottom row) of the timelike particles are plotted as a function of the density $\rho_s$ (left panels) and characteristic scale $r_s$ of the DM halo (right panels). \label{Fig.ISCO}}
\end{figure*}

To analyze the trajectory of a particle with mass $m$, total energy $E$, and angular momentum $L$ in the vicinity of a Schwarzschild-like BH in the DM halo, we start with the Hamilton-Jacobi equation \cite{Misner73,Shaymatov21pdu,Shaymatov22a}:
\begin{eqnarray}\label{eq.H-J}
    g^{\mu\nu}\frac{\partial S}{\partial x^\mu} \frac{\partial S}{\partial x^\mu}=-m^2\, ,
\end{eqnarray}
where $S=-Et+L\phi+S(r)_{r}+S_\theta(\theta)$ is the Hamilton-Jacobi action. Subsequently, we can express the Lagrangian of a test particle as:
\begin{eqnarray}\label{eq.Lagrangian}
    \mathcal{L}=\frac{1}{2}mg_{\mu\nu}u^{\mu}u^{\nu}\, ,
\end{eqnarray}
which enables us to find the four-momenta as:
\begin{eqnarray}\label{eq.momenta}
p_{\mu}=\frac{\partial\mathcal{L}}{\partial\Dot{x}}=mg_{\mu\nu}u^{\nu}\, .
\end{eqnarray}
Based on the above equations, we illustrate the trajectory of test particles starting from a point $r_0$ with the appropriate specific energy $\mathcal{E}=\frac{E}{m}$ and angular momentum $l=\frac{L}{m}$ \cite{Uktamov:2024ckf}, as shown in Fig~\ref{Fig.trajectory} \cite{Kolos:2023oii}. It can be observed that there are two types of orbits in Fig.~\ref{Fig.trajectory}: captured and bound orbits. An increase in the values of the DM halo parameters, $\rho_s$ and $r_s$, results in the orbits being captured.

For further analysis, we shall confine the motion to the equatorial plane (i.e., $\theta=\frac{\pi}{2}$):
\begin{subequations}\label{abcd}
\begin{align}
& \Dot{t}=-\frac{\mathcal{E}}{f(r)}\, ,
\\ 
& \Dot{\phi}=\frac{l}{r^2}\, ,
\\
& \Dot{r}^2=\mathcal{E}^2-V_{eff}(r)\, ,
\end{align}
\end{subequations}
where the effective potential, $V_{eff}(r)$, is given by 
\begin{eqnarray}\label{eq.effective}
 V_{eff}(r)&=&\Bigg[1-\frac{2M}{r}-\frac{2M_D(r)}{r_s}\left(1+\frac{r_s}{r}\right)\nonumber\\&\times&\log{\left(1+\frac{r_s}{r}\right)}\Bigg]\left(1+\frac{l^2}{r^2}\right)\, . 
\end{eqnarray}
We further examine the effective potential $V_{\rm eff}$, for the timelike particles around the BH in the DM halo characterized by the Dehnen-type density profile, highlighting how the presence of the novel Dehnen-type DM halo affects gravitational dynamics
and modifies particle trajectories. To accomplish this, we illustrate the radial behavior of the effective potential for various possible cases, as shown in Fig.~\ref{Fig.Veff}. As can be seen from Fig.~\ref{Fig.Veff}, it is evident that the effective potential has two extreme points that correspond to stable circular and unstable circular orbits. It can be observed from Fig.~\ref{Fig.Veff} that an increase in the density $\rho_s$ and the characteristic scale $r_s$ of the DM halo results in stable and unstable circular orbits of the particles expanding in the BH-DM halo system. It should also be noted that the effective potential curves exhibit a slight shift to the right, indicating larger radial distances ($r$), while the potential maximum decreases with increasing $\rho_s$ and $r_s$, as seen in Fig.~\ref{Fig.Veff}. 

We now turn to studying the innermost stable circular orbits (ISCOs). For the timelike particles to be on the ISCO, one needs to solve the following equations simultaneously \cite{Dadhich22a,Dadhich22IJMPD}
\begin{eqnarray}\label{eq.ISCO}
V_{eff}=\mathcal{E}^2_{ISCO},\,\,\frac{\partial V_{eff}}{\partial r}=\frac{\partial^2V_{eff}}{\partial r^2}=0\, .
\end{eqnarray}
\begin{figure}[ht!]
\hspace*{-1cm}
\includegraphics[width=0.5\textwidth]{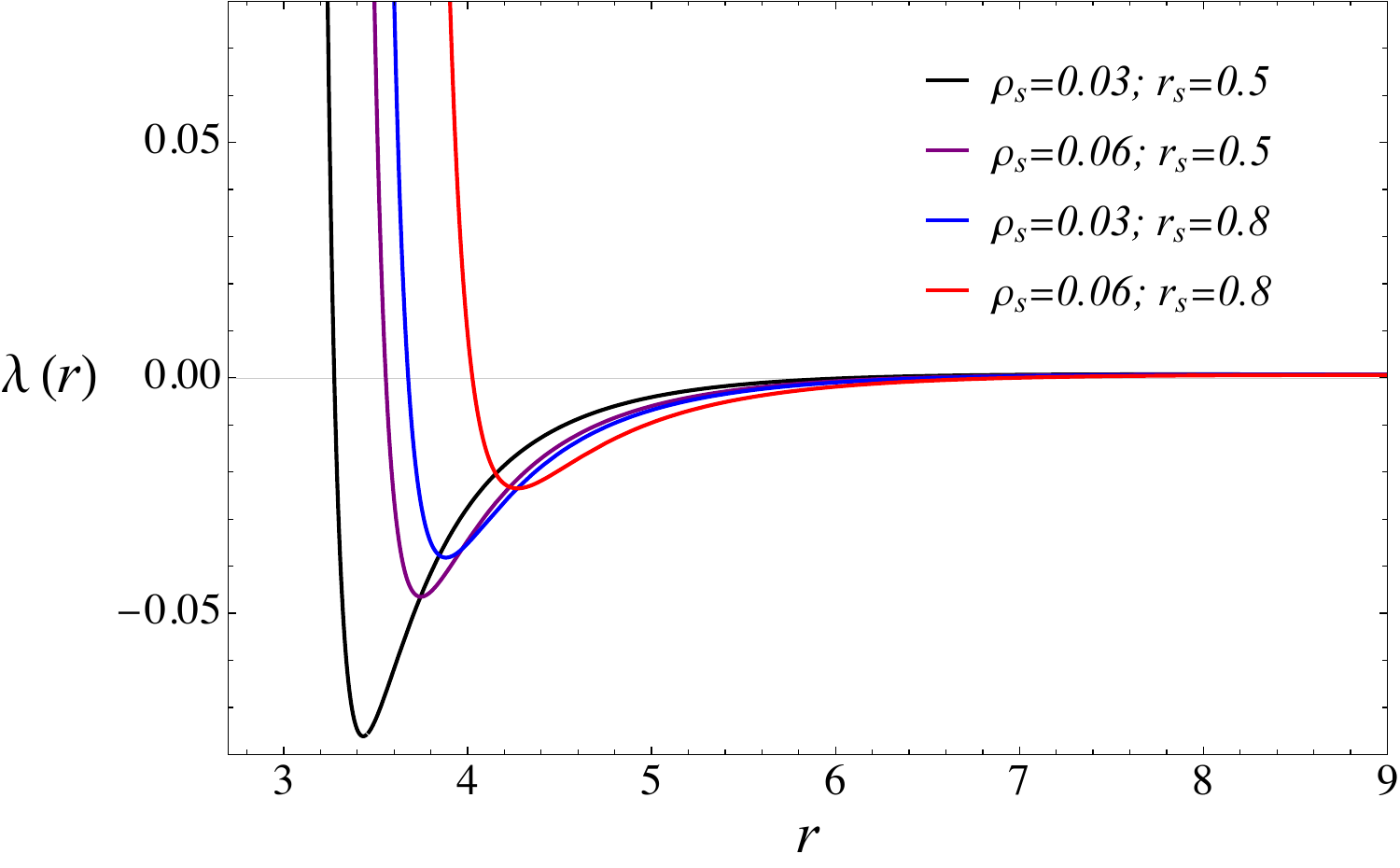}
\caption{Radial profile of the Lyapunov exponent for varying the density $\rho_s$ and characteristic scale $r_s$ of the DM halo.\label{Fig.Lyupanov}}
\end{figure}

We now proceed to analyze the behavior of the ISCO radii for particles orbiting around the Schwarzschild-like BH surrounded by the Dehnen-type DM halo. In Fig.~\ref{Fig.ISCO}, we highlight how the ISCO parameters, specifically the energy $\mathcal{E}_{ISCO}$ and the angular momentum $l_{ISCO}$ of the timelike particles, are influenced by the presence of the DM halo. As shown in Fig.~\ref{Fig.ISCO}, an increase in the DM halo $\rho_s$ leads to a corresponding increase in both $r_{ISCO}$ and $l_{ISCO}$ of timelike particles on the ISCO around the BH. Similarly, increasing the characteristic scale $r_s$ results in a larger radius $r_{ISCO}$; however, $l_{ISCO}$ initially decreases until a critical value of $r_s$ is reached, after which it begins to increase again, as depicted in Fig.~\ref{Fig.ISCO}.  

{Also, to be more informative we give numerical values of the $r_{ISCO}$ and $l_{ISCO}$ in Tables (\ref{Table ISCO 1}, \ref{Table ISCO 2}). One can conclude from these numerical values in Tables (\ref{Table ISCO 1}, \ref{Table ISCO 2}) that radius of the ISCO $r_{ISCO}$ increase monotonically with increasing DM halo parameters $r_s$ and $\rho_s$.}
\begin{table}[ht!]
    \centering
    \begin{tabular}{|l|c|c|r|}
     \hline
    $\rho_s$  & $r_s$ & $r_{ISCO}[M]$ & $l_{ISCO}[M]$ \\
    \hline
     0 & -   & 6   & $2\sqrt{3}$ \\
     0.03 & 0.1   & 6.00221   & 3.46539 \\
     0.03 & 0.2   & 6.01723   & 3.47427 \\
   0.03 & 0.3   & 6.05683   & 3.49797 \\
   0.03 & 0.4   & 6.13181   & 3.5434 \\
   0.03 & 0.5   & 6.25232   & 3.61723 \\
   0.03 & 0.6   & 6.42833   & 3.72604 \\
   0.03 & 0.8   & 6.98817   & 4.07541 \\
   \hline
    \end{tabular}
    \caption{{{The numerical values of the ISCO parameters $r_{ISCO}$ and $l_{ISCO}$ for different values of the characteristic scale of the DM halo $r_s$ at fixed $\rho_s$.}}}
    \label{Table ISCO 1}
\end{table}

\begin{table}[ht!]
    \centering
    \begin{tabular}{|l|c|c|r|}
     \hline
    $\rho_s$  & $r_s$ & $r_{ISCO}[M]$ & $l_{ISCO}[M]$ \\
    \hline
     0 & -   & 6   & $2\sqrt{3}$ \\
     0.01 & 0.4   & 6.04388   & 3.49051 \\
     0.02 & 0.4   & 6.08781   & 3.51695 \\
   0.03 & 0.4   & 6.13181   & 3.5434 \\
   0.04 & 0.4   & 6.17586   & 3.56987 \\
   0.05 & 0.4   & 6.21996   & 3.59636 \\
   0.06 & 0.4   & 6.26413   & 3.62287 \\
   0.07 & 0.4   & 6.30834   & 3.6494 \\
   \hline
    \end{tabular}
    \caption{{{The numerical values of the ISCO parameters $r_{ISCO}$ and $l_{ISCO}$ for different values of the characteristic density $\rho_s$ at fixed $r_s$.}}}
    \label{Table ISCO 2}
\end{table}

Finally, we turn to examining the Lyapunov exponent, which is a powerful tool for understanding the stability of timelike particles. Hereafter, we analyze the Lyapunov exponent to study the stability (instability) of circular orbits for timelike particles in the vicinity of the Schwarzschild-like BH in the DM halo \cite{2021EPJC...81..699R}:
\begin{eqnarray}\label{eq.Lyupanov}
     \lambda=\sqrt{-\frac{\partial_r^2V_{eff}}{2\Dot{t}^2}}\, .
 \end{eqnarray}
Additionally, we use requirements for circular orbits,  $\mathcal{E}^2=V_{eff}(r)$ and $\frac{\partial V_{eff}}{\partial r}=0$, which solve to give 
\begin{subequations}\label{eq.energy}
\begin{align}
& \mathcal{E}^2=\frac{2 f^2(r)}{2f(r)-rf'(r)}\, ,
\\ 
& l^2=\frac{r^3f'(r)}{2f(r)-rf'(r)}\, ,
\end{align}
\end{subequations}
where $'$ denotes a derivative with respect to $r$.

To better understand the Lyapunov exponents, we present the radial profile of $\lambda$ in Fig.~\ref{Fig.Lyupanov}. As shown in Fig.~\ref{Fig.Lyupanov}, the radii of stable ($\lambda<0$) and chaotic ($\lambda>0$) orbits of the timelke particles grow when the density $\rho_s$ and the characteristic scale $r_s$ of the DM halo are increased, coinciding with the analysis of the effective potential $V_{eff}(r)$, as depicted in Fig.~\ref{Fig.Veff}. In particular, bifurcation points occur when the curves intersect the horizontal axis (that is, $\lambda=0$), making the ISCO location. From Fig.~\ref{Fig.Lyupanov}, as a consequence of an increase in the DM halo parameters, $\rho_s$ and $r_s$, both the Lyapunov exponent curves and the corresponding stable circular orbits shift to the right to larger radial distances, as also reflected in Fig.~\ref{Fig.ISCO}.

{\section{Quasi-periodic oscillations models}}\label{Sec.V}

Now, we aim to find the epicyclic motion of the test particles orbiting around BHs in a Dehnen-type DM halo. {However,  microquasars are thought to host rapidly spinning BHs, and many QPO models rely critically on spin $a$. Therefore, we have to employ rotating version of the spacetime metric (\ref{eq.full-line}), which can be obtained by using Newman-Janis algorithm as (\cite{Uktamov:2025lsq})}:
{\begin{eqnarray}\label{eq.rotating line el.}
ds^2=g_{tt}dt^2+2g_{t\phi}dtd\phi+g_{\phi\phi}d\phi^2+g_{rr}dr^2+g_{\theta\theta}d\theta^2,
\end{eqnarray}
with metric components:
\begin{subequations}
\begin{eqnarray}
 g_{tt}&=&-\left[1-\frac{2Mr+2M_{D}r\log{(1+\frac{r_s}{r})^{\left(1+\frac{r}{r_s}\right)}}}{\Sigma}\right], \\ 
 g_{rr}&=&\frac{\Sigma}{\Delta},\\\nonumber
  g_{t\phi}&=&-a\sin^2{\theta}\left[\frac{2Mr+2M_{D}r\log\left(1+\frac{r_s}{r}\right)^{\left(1+\frac{r}{r_s}\right)}}{\Sigma}\right],\\\nonumber
g_{\phi\phi}&=&\sin^4{\theta}\Big[\frac{r^2+a^2}{\sin^2{\theta}}\nonumber\\&+&\frac{a^2\left(2Mr+2M_Dr\log(1+\frac{r_s}{r})^{\left(1+\frac{r}{r_s}\right)}\right)}{\Sigma}\Big],\\
g_{\theta\theta}&=&\Sigma,
\end{eqnarray}
\end{subequations}}

{\begin{subequations}\label{eq.metric components2}
\begin{align}
& \Sigma=r^2+a^2\cos^2{\theta},\\
& \Delta=r^2-2Mr+a^2-2M_Dr\log(1+\frac{r_s}{r})^{(1+\frac{r}{r_s})}.
\end{align}
\end{subequations}}
{To check whether obtained space-time metric for rotating BH in DM halo (\ref{eq.rotating line el.}) satisfies Einstein field equation is primary task. We have shown that obtained space-time metric for rotating BH in DM halo satisfies Einstein field equations in Appendix \ref{Sec.App}.}  

{Radial equation of (\ref{eq.H-J}) with the four-velocity $\Dot{x}^\mu=u^t(1,0,0,\Omega)$ can be expressed as $g_{tt,r}+2\Omega g_{t\phi,r}+\Omega^2g_{\phi\phi,r}=0$, which paves a way to express Keplerian frequency $\Omega$ as:
\begin{eqnarray}\label{eq.Keplerian}
    \Omega=-\frac{g_{t\phi,r}}{g_{\phi\phi,r}}\pm\sqrt{(\frac{g_{t\phi,r}}{g_{\phi\phi,r}})^2-\frac{g_{tt,r}}{g_{\phi\phi,r}}}\,,
\end{eqnarray}
considering Eqs.(\ref{eq.rotating line el.}),
\begin{eqnarray}
    \Omega = \frac{1}{a+r\sqrt{\frac{r}{M+M_D}}}\,,
\end{eqnarray}
which converted into Keplerian frequency for Kerr BH in the limit $\rho_s\to0$ ($M_D=0$).}

The epicyclic frequency is the oscillation of particles with radial $\Omega_r$ and vertical $\Omega_\theta$ frequencies around stationary points ($r_0,\,\,\theta_0$). The calculation of epicyclic frequencies involves a perturbative analysis of a stable circular orbit, introducing small displacements in both the radial $\delta r$ and tangential $\delta \theta$ coordinates, which gives harmonic oscillator equations as (\cite{2025PDU....4701805X,2022Univ....8..507T}):
\begin{subequations}\label{eq.oscillator equations}
\begin{align}
& \frac{d^2}{dt^2}\delta r+\Omega_r^2\delta r=0\,,\\
& \frac{d^2}{dt^2}\delta\theta+\Omega_\theta^2\delta\theta=0\,. 
\end{align}
\end{subequations}

Subsequently, these frequencies can be calculated as:
\begin{subequations}\label{eq.frequencies}
   \begin{align}
       & \Omega_r^2=\frac{1}{2g_{rr}\Dot{t}^2}\partial_r^2V_{eff}\,,\\
       &\Omega_\theta^2=\frac{1}{2g_{\theta\theta}\Dot{t}^2}\partial_\theta^2V_{eff}\,,
   \end{align} 
\end{subequations}
then after implying Eq.(\ref{eq.rotating line el.}) frequencies can be expressed as:
{\begin{subequations}\label{eq.frequencies2}
    \begin{align}
        & \Omega_r=\Omega\sqrt{\Psi},\\
        & \Omega_\theta=\Omega\sqrt{\Xi},
    \end{align}
\end{subequations}
where new variables are given in Appendix \ref{Sec.App}.}

{Following, Eqs. (\ref{eq.frequencies2}) should be multiplied by the factor $\frac{c^3}{GM}$ to convert frequencies in physical units into SI (Hz) unity:
\begin{eqnarray}\label{eq.Hz}
    \nu_i=\frac{1}{2\pi}\frac{c^3}{GM}\Omega_i (\text{Hz})\,\,\,i=(r,\theta,\phi)\,. 
\end{eqnarray}}

\subsection{QPO studies in RP1 model: MCMC analysis}
\begin{figure*}[ht!]
\centering
\includegraphics[width=0.45\textwidth]{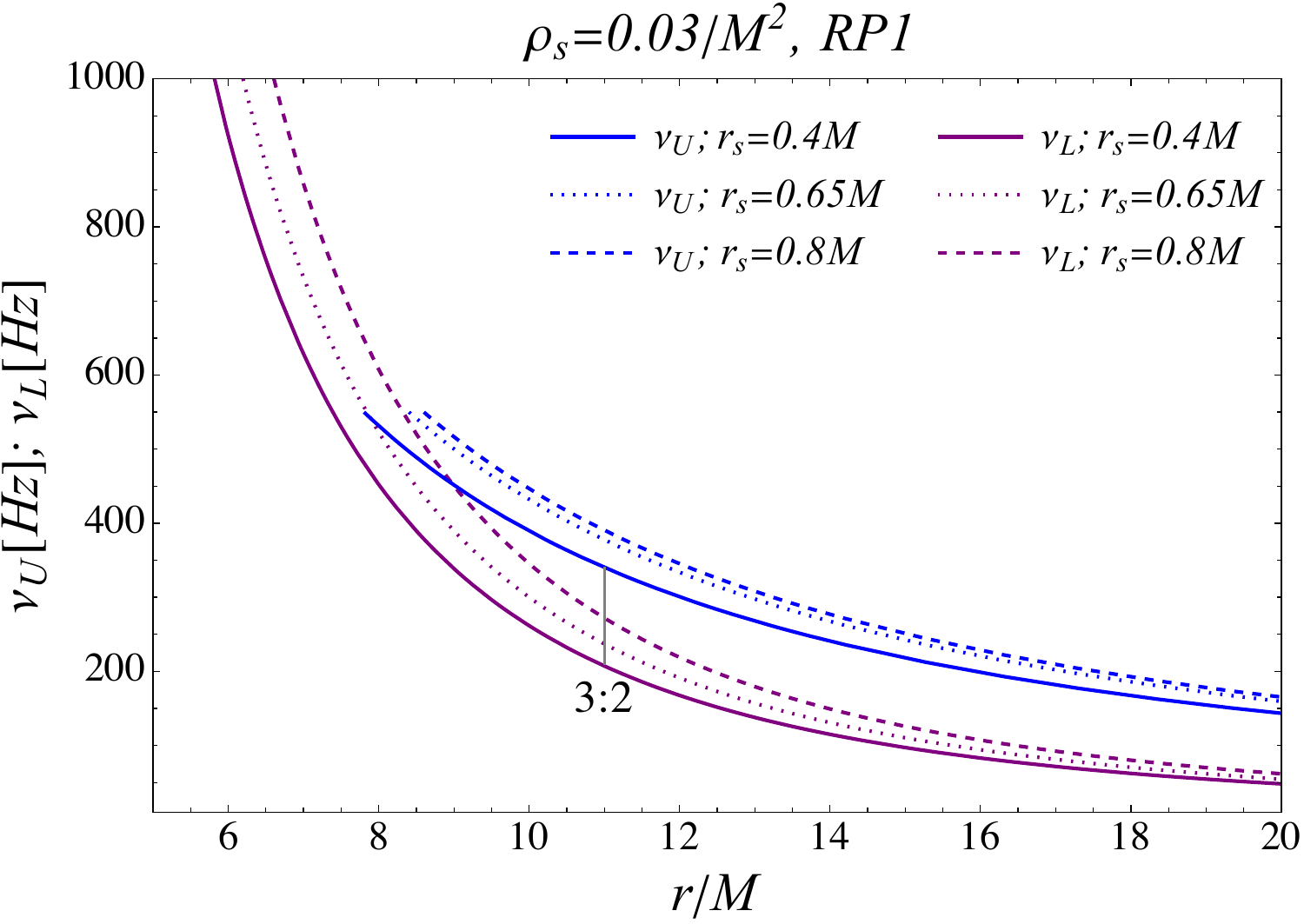}
\includegraphics[width=0.45\textwidth]{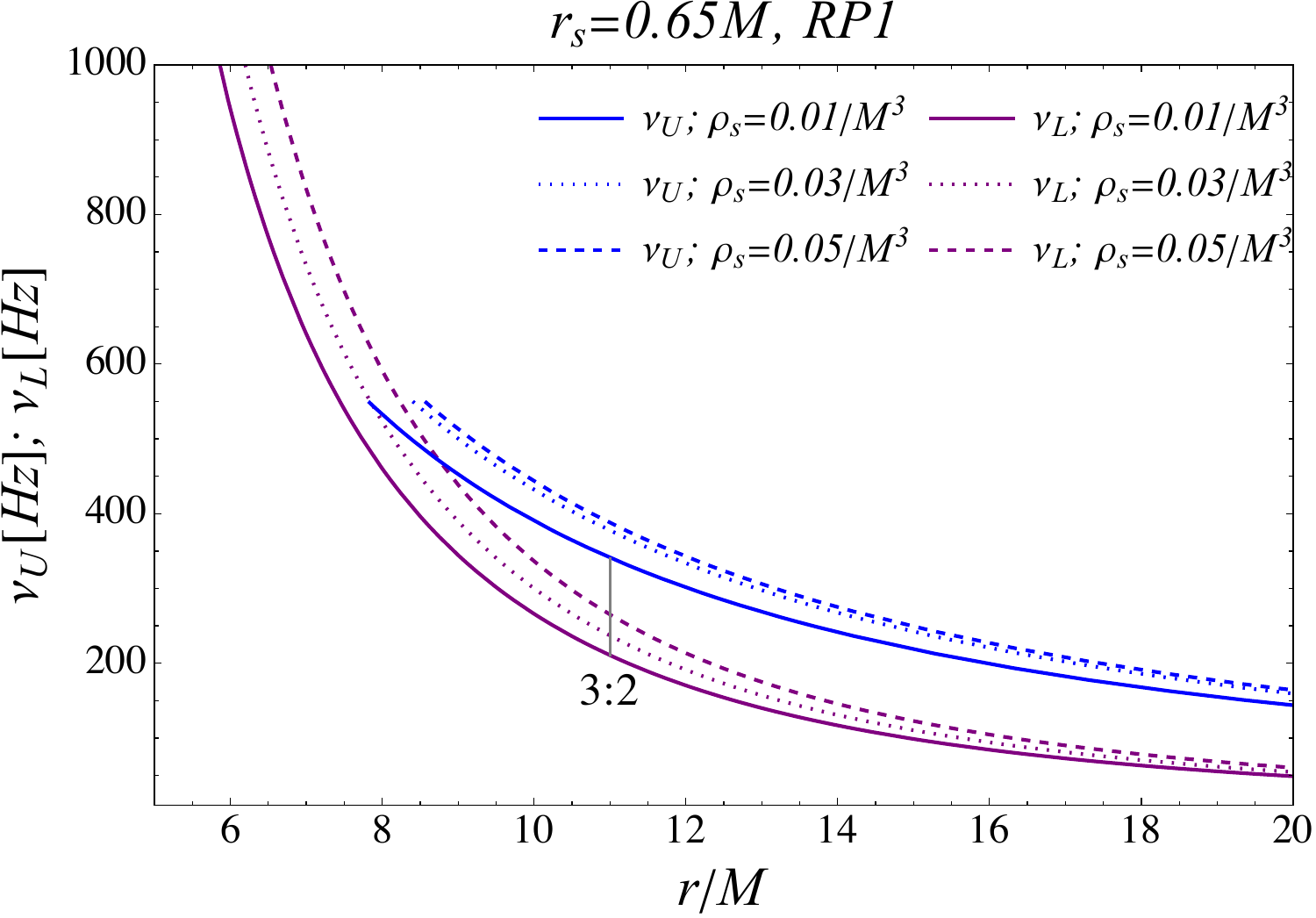}
\caption{{The functional relationship between the QPO frequencies (upper $\nu_U$ and lower $\nu_L$) and the  $r$. This dependence is illustrated for varying values of the parameter $\rho_s$ (right panel) and $r_s$ (left panel). The radii where the frequency ratio $\nu_U/\nu_L=3/2$ are also shown. Here we set BH spin as $a=0.7M$.\label{Fig.QPO}}}
\end{figure*}

The strong gravity of compact objects warps spacetime and influences nearby matter, which in turn produces electromagnetic radiation. The electromagnetic radiation from the accretion disks often exhibits quasiperiodic oscillations (QPOs). These QPOs are categorized by their often labelled as an upper ($\nu_U$) and lower ($\nu_L$) frequency pair. {One promising model, the  relativistic precession RP1 model, has the upper frequency as the orbital frequency ($\nu_U = \nu_\theta$), and the lower frequency is the difference between Keplerian frequency and the radial frequency  ($\nu_L = \nu_\phi-\nu_r$) \cite{Kolos:2020ykz,1999ApJ...524L..63S} Also, we have plotted radial dependence of the upper $\nu_U$ frequency and lower frequency $\nu_L$ in Fig.(\ref{Fig.QPO}) for different values of the DM parameters $\rho_s$ and $r_s$. Hence, we will employ RP1 model (\cite{1999ApJ...524L..63S,Kolos:2020ykz}) to get constrain for DM halo parameters $\rho_s$, $r_s$ via MCMC analysis.} Then, in Table \ref{Table 1} we have given the observed range of the mass, spin, upper and lower QPO frequencies.

\begin{table}[ht!]
    \centering
    \resizebox{.5\textwidth}{!}{
    \begin{tabular}{|l|c|c|c|c|r|}
     \hline
        & GRO J1655-40 & GRS 1915+105 & XTE J1550-564 \\
    \hline
      $M/M_0$   & $5.7-5.1$    & $10-18$ & $8.4-10.8$\\
      $\nu_U(Hz)$  & $439-443$   & $165-171$ & $273-279$\\
      $\nu_L(Hz)$   & $294-302$   & $108-118$ & $179-189$\\
      $a[M]$   & $0.65-0.75$   & $0.96-0.99$ & $0.31-0.34$\\
   \hline
    \end{tabular}
    }
    \caption{{The selected X-ray binary systems' characteristics, including their mass estimates and the corresponding upper and lower quasiperiodic oscillation frequencies.\cite{2025PDU....4701805X,2006ARA&A..44...49R,Shafee:2005ef,Miller:2013rca,Franchini:2017xzu}}}
    \label{Table 1}
    \end{table}

Now using observed data of three X-ray binary systems, GRO J1655-40, GRS 1915+105, XTE 1550-564 in Table.\ref{Table 1}, we will employ MCMC analysis to constrain DM halo parameters $\rho_s$ and $r_s$. Posterior distribution can be expressed as:
\begin{eqnarray}\label{eq.post.dist}
    P(\Theta|\mathcal{D},\mathcal{M})=\frac{\mathcal{L}(\Theta|\mathcal{D},\mathcal{M})\pi(\Theta|\mathcal{M})}{P(\mathcal{D}|\mathcal{M})},
\end{eqnarray}
here, $\mathcal{L}(\Theta|\mathcal{D},\mathcal{M})$ represents the probability of the data D under model M for a specific parameter set $\Theta$, $P(\mathcal{D}|\mathcal{M})$ is the marginal likelihood, and $\pi(\Theta|\mathcal{M})$ is the prior distribution. In our analysis, we also set priors as Gaussian priors (Table \ref{Table 2}), with boundary conditions:
\begin{eqnarray}\label{eq.Gaussian}
\pi(\Theta_i)\sim\exp{\left[\left(\frac{\Theta_i-\Theta_{0,i}}{\sigma_i}\right)^2\right]},
\end{eqnarray}
with standart deviation $\sigma_i$.
\begin{table}[h!]
\centering
\resizebox{.5\textwidth}{!}{
\begin{tabular}{|c|cc|cc|cc|cc|cc|}
\hline
& \multicolumn{2}{c|}{GRO J1655-40} & \multicolumn{2}{c|}{GRS 1915+105} & \multicolumn{2}{c|}{XTE J1550-564}  \\
 & $\mu$ & $\sigma$ & $\mu$ & $\sigma$ & $\mu$ & $\sigma$ \\
\hline
$M \ (M_\odot)$ & 5.4 & 0.1 & 14.1 & 1.09 & 9.6 & 0.3  \\
$r(M)$ & 12.1 & 0.5 & 12.5 & 0.5 & 9.1 & 0.3  \\
$\rho_s(1/10^{19}M_\odot^2)$ & 0.03 & 0.009 & 0.05 & 0.003 & 0.011 & 0.0015 \\
$r_s(10^{17}M_\odot)$ & 0.6 & 0.09 & 0.6 & 0.07 & 0.6 & 0.02 \\
$a(M)$ & 0.7 & 0.05 & 0.97 & 0.01 & 0.32 & 0.01 \\
\hline
\end{tabular}
}
\caption{{Gaussian prior on   BH in DM halo mass from QPO constraints.}}
\label{Table 2}
\end{table}
The likelihood function $\mathcal{L}$ can be written as the sum of the upper likelihood function $\mathcal{L}_{up}$ and the lower likelihood function $\mathcal{L}_{low.}$:
\begin{eqnarray}\label{eq.likelihood}
\mathcal{L}=\mathcal{L}_{up}+\mathcal{L}_{low.}=&-\frac{1}{2}\Sigma_i\frac{(\nu^i_{up,obs.}-\nu^i_{up,th})^2}{(\sigma^i_{up,obs.})^2}-\\\nonumber
&-\frac{1}{2}\Sigma_i\frac{(\nu^i_{low.,obs.}-\nu^i_{low.,th})^2}{(\sigma^i_{low.,obs.})^2},
\end{eqnarray}
where the terms $\nu_{up,obs.}$ and $\nu_{low.,obs.}$ represent the observed upper and lower QPO frequencies, while $\nu_{up,th.}$ and $\nu_{low.,th.}$ denote the corresponding theoretical predictions for these frequencies generated by the model.
\begin{figure*}
    \centering
\includegraphics[scale=0.25]{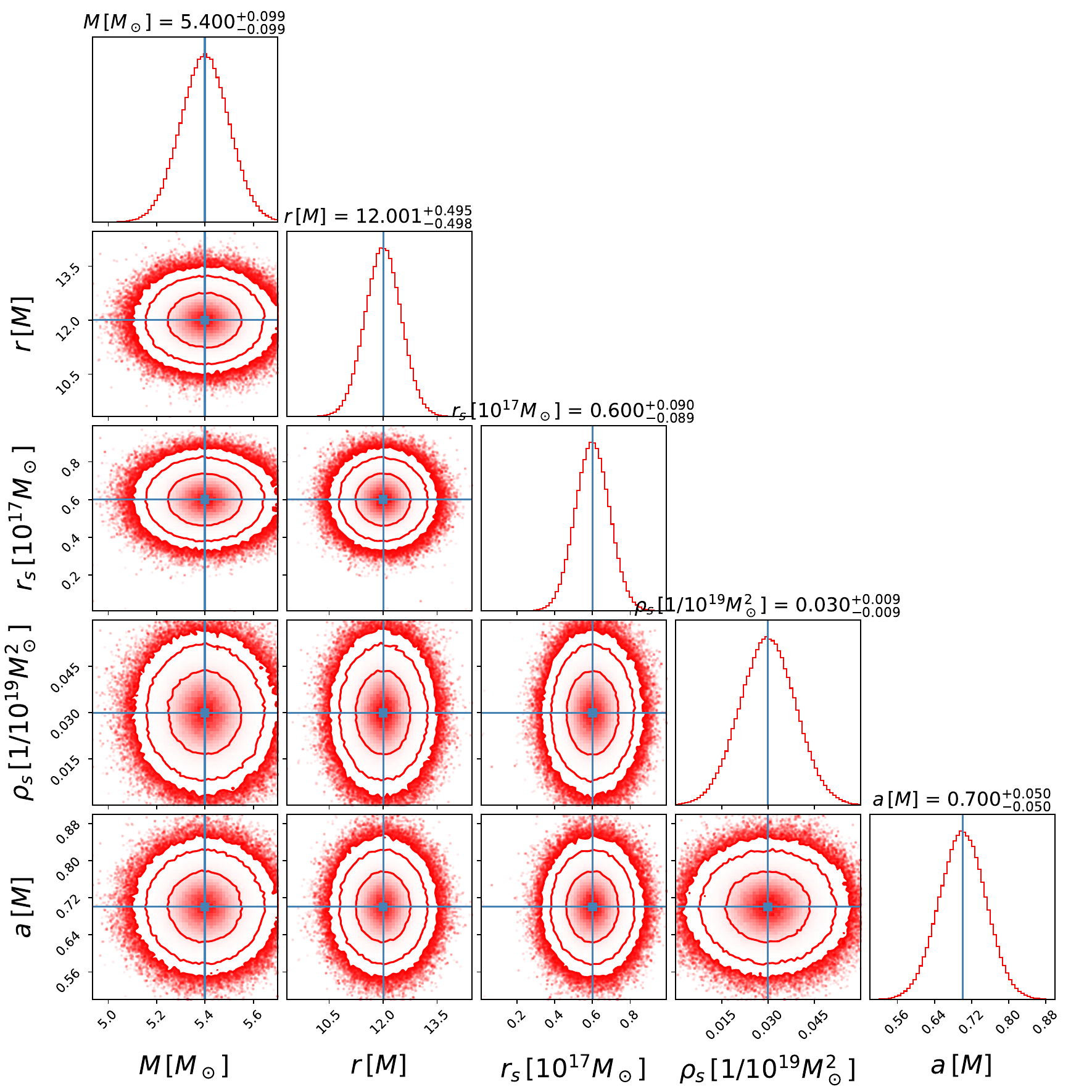}
\includegraphics[scale=0.25]{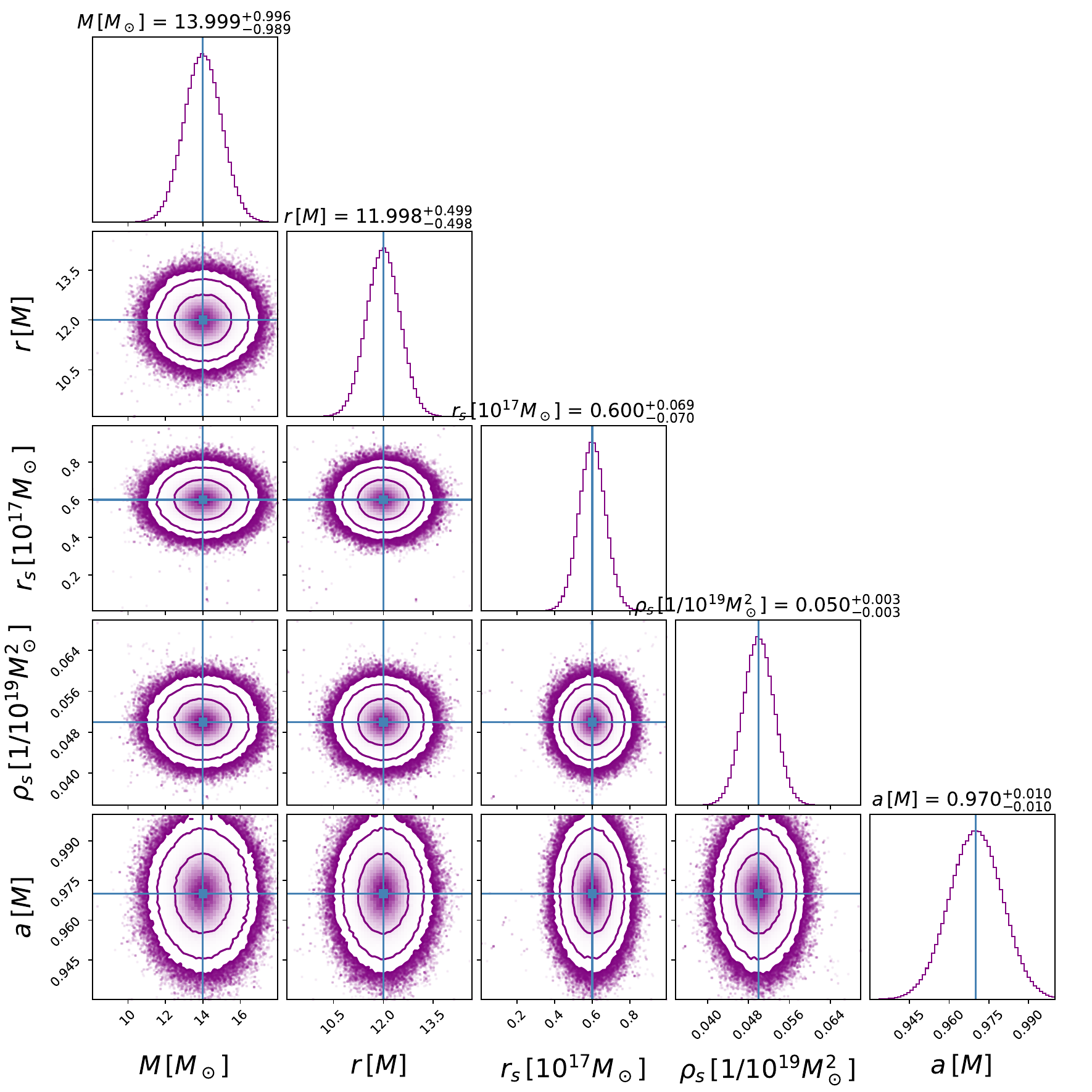}
\includegraphics[scale=0.25]{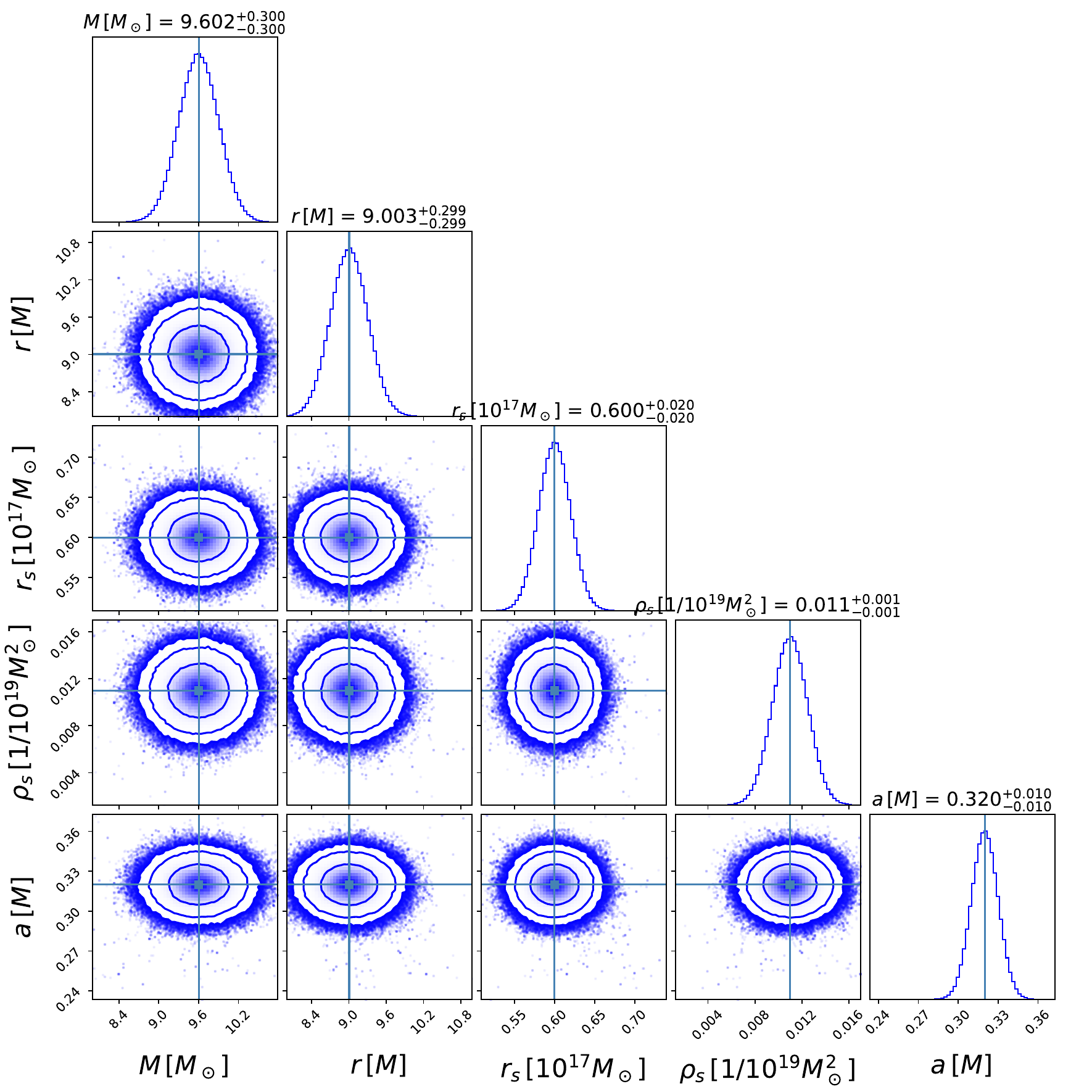}
\caption{{Correlations between twin-peak QPO frequencies in the RP1 model for a  BH with a Dehnen-type DM halo (upper left one for "GRO J1655-40", upper right one for "GRS 1915+105", lower one for "XTE J1550-564").}}
    \label{fig:MCMC}
\end{figure*}

Subsequently, using the emcee package, we have employed MCMC analysis to constrain DM halo parameters $\rho_s$ and $r_s$ in Fig.\ref{fig:MCMC}. The figure constitutes three panels: upper left one for "GRO J1655-40", upper right one for "GRS 1915+105", lower one for "XTE J1550-564" and contours in these graphs (Fig.\ref{fig:MCMC}) show $1\sigma$ (68$\%$), $2\sigma$ (90$\%$), and $3\sigma$ (95$\%$) confidence regions for the posterior distributions. To be more informative, we have given the best-fit parameters of the Schwarzchild-like BH immersed in Dehnen-type DM halo in the Table.\ref{Table 3}

{We can see our obtained best-fit results well coincide with previous results for DM halo parameters (\cite{2020MNRAS.494.4291C,1996ApJ...462..563N}) as:
\begin{eqnarray}\label{eq.comparision}
  \rho_s\sim \frac{10^{-2}}{(10^{19}M_\odot)^2} \frac{c^6}{G^3}\sim10^{-24}g/cm^{3}\,,\\\nonumber
  r_s\sim10^{17}M_\odot\frac{G}{c^2}\sim 10^{22}cm\sim 10\text{kpc}\,,
\end{eqnarray}
which indicate physically reasonability of the our model.}

\begin{table}[ht!]
    \centering
    \resizebox{.47\textwidth}{!}{
    \begin{tabular}{|l|c|c|c|r|}
     \hline
        & GRO J1655-40 & GRS 1915+105 & XTE J1550-564 \\
    \hline
      $M(M_\odot)$   & $5.4^{+0.099}_{-0.099}$    & $13.997^{+0.994}_{-0.995}$ & $9.599^{+0.300}_{-0.298}$\\
      & & &\\
      $r(M)$  & $12.001^{+0.494}_{-0.499}$   & $12.002^{+0.498}_{-0.499}$ & $9.002^{+0.298}_{-0.299}$\\
      & & &\\
      $r_s(10^{17}M_\odot)$   & $0.6^{+0.090}_{-0.089}$   & $0.6^{+0.069}_{-0.069}$ & $0.6^{+0.020}_{-0.020}$\\
      & & &\\
      $\rho_s(1/10^{19}M_\odot^2)$   & $0.03^{+0.009}_{-0.009}$   & $0.05^{+0.003}_{-0.003}$ & $0.011^{+0.001}_{-0.001}$\\
      & & &\\
      $a(1/M_\odot)$   & $0.7^{+0.05}_{-0.05}$   & $0.97^{+0.01}_{-0.01}$ & $0.32^{+0.01}_{-0.01 }$\\
   \hline
    \end{tabular}}
    \caption{{{Best-fit BH in DM halo parameters derived from QPO observations of the selected X-ray sources.}}}
    \label{Table 3}
\end{table}

\section{Conclusion}\label{Sec:conclusion}

Black holes interact gravitationally with their surrounding environments, making the study of BH-DM systems an increasingly important task in astrophysical contexts. It is widely believed that supermassive BHs at the centers of galaxies are embedded in complex, dynamic matter distributions, including DM halos \cite{Rees84ARAA,Kormendy95ARAA,Iocco15NatPhy,Bertone18Nature,Valluri04ApJ,Akiyama19L1}. Motivated by this, we derived a novel analytical Schwarzschild-like BH solution, representing a static BH surrounded by a DM halo characterized by a Dehnen-type density profile $(1,4,2)$. 

We investigated the properties of this BH by analyzing the spacetime curvature properties. Specifically, we computed the curvature invariants of the spacetime metric to determine the presence of singularities. Our results indicated that a spacetime singularity appears at $r=0$, as illustrated in Fig.~\ref{Fig.Curvature}. We also examined the energy conditions, which are fundamental characteristics of the physical viability of spacetime. Our analysis showed that all energy conditions are well satisfied for the derived BH solution within the Dehnen-type DM halo. This conclusion was further supported by graphical results, as presented in Fig.~\ref{Fig.EC}, with analytical and graphical analyzes consistently confirming the validity of all energy conditions.

We also investigated the timelike geodesics of particles in the BH-DM spacetime, highlighting how the novel Dehnen-type DM halo influences the gravitational dynamics and affects particle trajectories. We analyzed the trajectory of timelike particles with the appropriate specific energy $\mathcal{E}$ and angular momentum $l=\frac{L}{m}$. We showed that there exist two types of orbits: captured and bound orbits. Our analysis indicated that these orbits become captured with increasing $\rho_s$ and $r_s$, as represented in Fig.~\ref{Fig.trajectory}. Using the Lagrangian formalism, we derived the effective potential and analyzed its radial behavior for timelike particle motion. The effective potential exhibits two extrema, corresponding to stable and unstable circular orbits. Our results showed that increasing the DM halo's density $\rho_s$ and characteristic scale $r_s$ leads to an outward shift of both stable and unstable circular orbits, effectively expanding the influence region of the BH–DM system. The effective potential curves exhibit a slight shift to the right, indicating larger orbital radii, while the maximum effective potential decreases with an increase in $\rho_s$ and $r_s$, as illustrated in Fig.~\ref{Fig.Veff}.

We also examined the influence of the DM halo on the ISCO parameters, specifically, the energy $\mathcal{E}_{ISCO}$, angular momentum $l_{ISCO}$ and the ISCO radius $r_{ISCO}$ for timelike particles orbiting the Schwarzschild-like BH surrounded by a Dehnen-type DM halo. Our analysis showed that increasing the DM density $\rho_s$ results in a corresponding increase in both $r_{ISCO}$ and $l_{ISCO}$. Interestingly, the characteristic scale $r_s$ has a nonmonotonic effect on $l_{ISCO}$, i.e, while $r_{ISCO}$ increases with $r_s$, the angular momentum $l_{ISCO}$ initially decreases up to a critical value of $r_s$, beyond which it begins to increase again, as shown in Fig.~\ref{Fig.ISCO}. Finally, we investigated the Lyapunov exponent to understand the stability of timelike particle orbits in the BH vicinity. The radial profile of the Lyapunov exponent, presented in Fig.~\ref{Fig.Lyupanov}, clearly shows a slight rightward shift in the curves with increasing $\rho_s$ and $r_s$, suggesting that the corresponding stable circular orbits are displaced to larger radial distances. This behavior is consistent with the change in ISCO radii observed in Fig.\ref{Fig.ISCO}. 

We further performed an MCMC analysis using the emcee package to constrain the parameters of the rotating BH in Dehnen-type DM halo. The results, presented in Fig. \ref{fig:MCMC} and Table \ref{Table 3} for the microquasars GRO J1655-40, GRS 1915+105, and XTE J1550-564, reveal distinct halo properties for each source. Specifically, the QPOs from GRS 1915+105 indicate the highest DM halo density, while those from GR) J1655-40 correspond to the halo with the largest core radius.

Our approach offers fundamental astrophysical implications, particularly concerning the significant role of DM halos around supermassive BHs in galactic nuclei. Specifically, the Dehnen-type density distribution used here supports the existence of a BH solution with a DM halo within the density profile $(1,4,2)$. Our findings could offer a novel perspective on the formation of DM halos, providing a potential alternative framework for understanding their fundamental properties.

\section*{Acknowledgements}

S.S. is supported by the National Natural Science Foundation of China under Grant No. W2433018.

\section{Appendix}\label{Sec.App}

\begin{figure*}
    \centering
\includegraphics[scale=0.3]{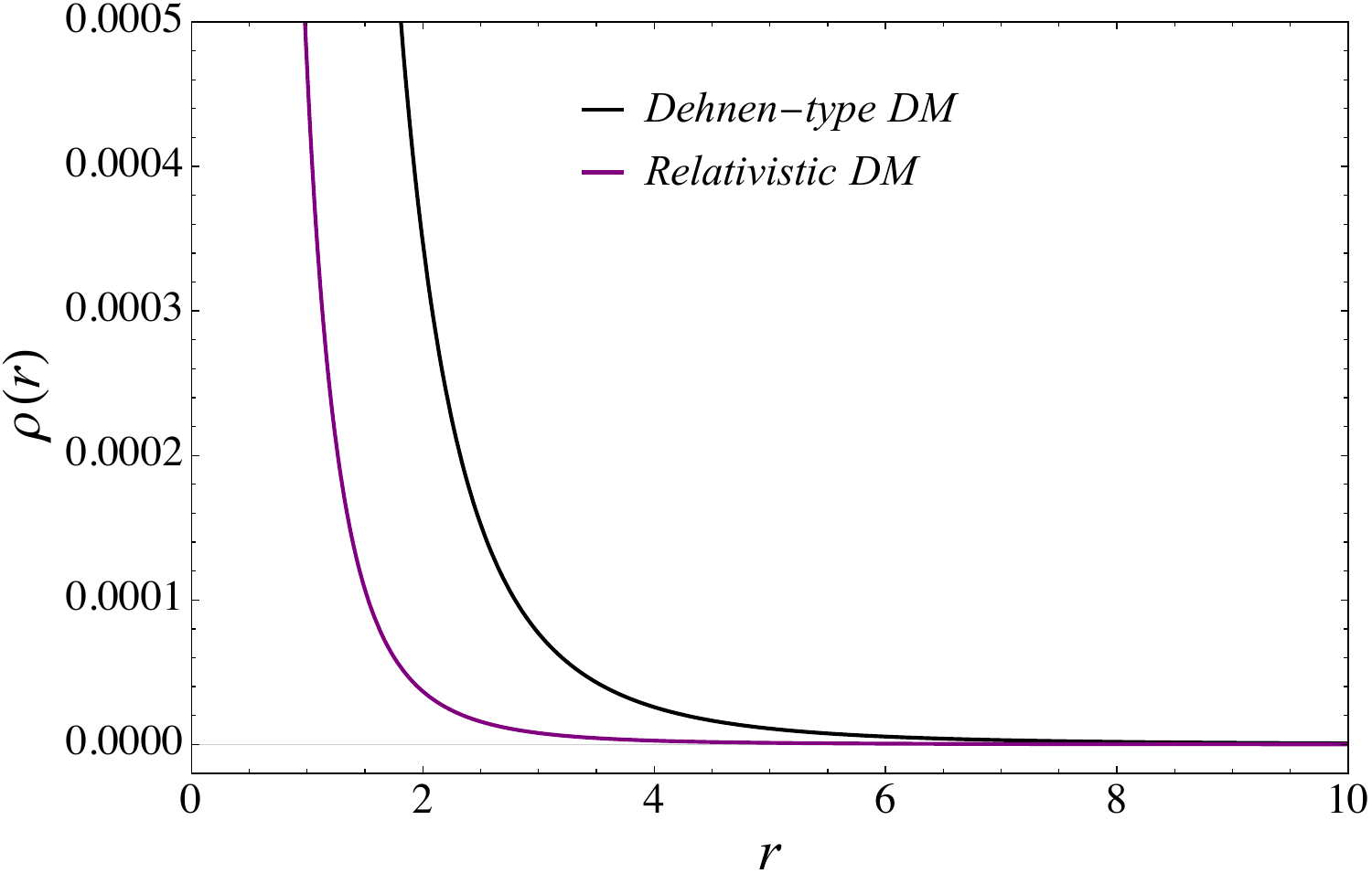}
\caption{{The radial dependence of the density distribution for Dehnen-type DM halo for both cases considering relativistic effects (purple solid lines) and do not considering relativistic effects (Black solid lines). Here we take DM halo parameters as $\rho_s=0.005$ and $r_s=0.4$.}}
    \label{fig:Density}
\end{figure*}

{The $G_{tt}$ component of the Einstein field equations for the obtained spacetime metric (\ref{eq.full-line}) has the following form:}
{\begin{eqnarray}
 G_{tt}&=&\frac{8\pi\rho_sr_s^4}{r^2(r+r_s)^2\Big[1-\frac{2M}{r}-8\pi\rho_sr_s^2\log{(1+\frac{r_s}{r})}\Big]}\\\nonumber
    &\times&\Big[1+\frac{r}{r_s}-(1+\frac{r}{r_s})^2\log{(1+\frac{r_s}{r})}\Big]\,, 
\end{eqnarray}
which is fully consistent with energy density given in (\ref{eq.elements of the energy-mom.}). However, we can see the  density distribution of the obtained space time (\ref{eq.elements of the energy-mom.}) has extra term  from  the Dehnen-type DM halo density distribution as:
\begin{eqnarray}\label{eq.extra}
\rho(r)=\rho_D\Big[1+\frac{r}{r_s}-(1+\frac{r}{r_s})^2\log{(1+\frac{r_s}{r})}\Big], 
\end{eqnarray}
here $\rho_D$ is the Dehnen-type density distribution (\ref{eq.density}). The reason behind this difference is the relativistic effects of the BH, such as gravitational interaction energy between the black hole and the dark matter halo. To be more informative we give radial dependence of the both density distributions in Fig.(\ref{fig:Density}). One can see from Fig.(\ref{fig:Density}) that for astrophysical considerable values of the $r$  the effect of the relativity is negligable.}

{\subsection{Checking Einstein field equations for rotating BH in DM halo}}
{The nonzero  components of the Einstein tensor for the line element of the rotating BH in DM halo can be expressed as (\ref{eq.rotating line el.}):}
{\begin{eqnarray}\label{Eq.Einstein comp.}
    G_{tt}=-\frac{4F^2+2r\left[r^2+a^2\left(2-\cos^2{\theta}\right)\right]F_{,r}}{\Sigma^3}
    \end{eqnarray}
    \begin{eqnarray}\nonumber
\frac{-2F\left[r^2+a^2\left(2-\cos^2{\theta}\right)+2rF_{,r}\right]-a^2\sin^2{\theta}\Sigma F_{,rr}}{\Sigma^3}, 
\end{eqnarray}
\begin{eqnarray}
    G_{rr}=-\frac{2\left(F-rF_{,r}\right)}{\Sigma\Delta}\,,
\end{eqnarray}
\begin{eqnarray}
G_{\theta\theta}=\frac{2\left(F-rF_{,r}\right)+\Sigma F_{,rr}}{\Sigma}\,,
\end{eqnarray}
\begin{eqnarray}
    G_{t\phi}=-\frac{a\sin^2{\theta}\Big[4F\left(a^2+r^2+rF_{,r}\right)-4F^2-}{\Sigma^3}\\\nonumber
    \frac{-\left(a^2+r^2\right)\left(4rF_{,r}-\Sigma^3F_{,rr}\right)\Big]}{\Sigma^3}\,,
\end{eqnarray}
\begin{eqnarray}\label{eq.Ein.2}
    G_{\phi\phi}=-\frac{\sin^2{\theta}}{\Sigma^3}\{4a^2\sin^2{\theta}F^2-F\Big[2(a^2+r^2)\\\nonumber
(r^2+a^2[2-\cos^2{\theta}])+4a^2r\sin^2{\theta}F_{,r}\Big]+(a^2+r^2)\\\nonumber
\Big[2rF_{,r}(r^2++a^2(2-\cos^2{\theta}))-(a^2+r^2)\Sigma F_{,rr}\Big]\}
\end{eqnarray}}
{here we have introduced a new variable as $F(r)=\frac{r^2\left[1-f(r)\right]}{2}$ . The appropriate  orthogonal bases for space-time metric (\ref{eq.rotating line el.}) can be expressed as (\cite{Azreg-Ainou14PRD},\cite{Jusufi:2019nrn}):
\begin{subequations}\label{eq.tetrads}
    \begin{align}
        &e_t^\mu=-\frac{\left(r^2+a^2,0,0,a\right)}{\sqrt{\Sigma\Delta}}\,,\,\,e_r^\mu=-\frac{\sqrt{\Delta}\left(0,1,0,0\right)}{\sqrt{\Sigma}}\,,\\
        &e_\theta^\mu=-\frac{\left(0,0,1,0\right)}{\sqrt{\Sigma}}\,,\,\,e_\phi^\mu=\frac{\left(a\sin^2{\theta},0,0,1\right)}{\sqrt{\Sigma}\sin{\theta}}\,.
    \end{align}
\end{subequations}}
{
Then $\rho$ component of the energy momentum tensor $T_{\mu}^{\nu}=\text{diag}[-\rho,p_r,p_\theta,p_\phi]$ can be expresses as:
\begin{eqnarray}
    -\rho=\frac{1}{8\pi}e_{t}^\mu e_{t}^\nu G_{\mu\nu}\,,
\end{eqnarray}
or considering Eqs.:(\ref{Eq.Einstein comp.}-\ref{eq.Ein.2})
\begin{eqnarray}\label{eq.rho final}
    \rho(r)=\frac{2(rF_{,r}-F)}{4\pi\Sigma^2}=
\end{eqnarray}
\begin{eqnarray}\nonumber
    \frac{\rho_D}{(1+(\frac{a}{r})^2\cos^2{\theta})^2}\left[1+\frac{r}{r_s}-(1+\frac{r}{r_s})^2\log(1+\frac{r_s}{r})\right]\,,
\end{eqnarray}
which turns into Eq.(\ref{eq.extra}) in the limit $a\to0$.}

{\subsection{Variables for frequencies in Eqs.(\ref{eq.frequencies2})}
\begin{eqnarray}\nonumber
    \Psi&=&1-\frac{3a^2}{r^2}+\frac{4(\mathcal{M}-M_D)}{r}-\frac{6(M+\mathcal{M})}{r}+\frac{8a\sqrt{M+M_D}}{r\sqrt{r}}+\\\nonumber
    &+&\frac{M_Dr_s\Big[a^2+r(r-2M)-2\mathcal{M}r\Big]}{r^2(r+r_s)(M+M_D)}\,,
\end{eqnarray}
\begin{eqnarray}\nonumber
    \Xi=1-\frac{4a(M+\mathcal{M})}{r\sqrt{r(M+M_D)}}+\frac{a^2(3M+M_D+2\mathcal{M})}{r^2(M+M_D)}\,,
\end{eqnarray}
here $\mathcal{M}=M_D(1+\frac{r}{r_s})\log{(1+\frac{r_s}{r})}$.}

\bibliographystyle{spphys}
\bibliography{Refs}
\end{document}